\documentclass[aps,showpacs,twocolumn,prd,superscriptaddress]{revtex4}

\usepackage{amsmath}
\usepackage{latexsym}
\usepackage{graphicx}
\usepackage{color}
\usepackage{enumerate}

\newcommand{\beq}{\begin{equation}}
\newcommand{\eeq}{\end{equation}}
\newcommand{\bea}{\begin{eqnarray}}
\newcommand{\eea}{\end{eqnarray}}
\newcommand{\ba}{\begin{array}}
\newcommand{\ea}{\end{array}}
\newcommand{\MSun}{{\rm M}_{\odot}}
\newcommand{\X}{\bf X}
\newcommand{\A}{\bf A}

\newcommand{\bA}{\bar{\bf A}}
\newcommand{\Abar}{\bar{A}}
\newcommand{\Ebar}{\bar{E}}
\newcommand{\Tbar}{\bar{T}}
\newcommand{\balpha}{\boldsymbol \alpha}

\newcommand{\hB}{h_B}
\newcommand{\ie}{i.~e.~}

\def\leq{\,\raise 0.4ex\hbox{$<$}\kern -0.8em\lower 0.62ex\hbox{$-$}\,}
\def\geq{\,\raise 0.4ex\hbox{$>$}\kern -0.8em\lower 0.62ex\hbox{$-$}\,}
\def\pm{\,\raise 0.4ex\hbox{$+$}\kern -0.8em\lower 0.62ex\hbox{$-$}\,}

\begin{document}

\title{Impact of mergers on LISA parameter estimation
for nonspinning black hole binaries }

\author{Sean T. McWilliams}
\email{Sean.T.McWilliams@nasa.gov}
\affiliation{Gravitational Astrophysics Laboratory, NASA Goddard Space Flight Center, 8800 Greenbelt Rd., Greenbelt, MD 20771, USA}
\author{James Ira Thorpe}
\affiliation{Gravitational Astrophysics Laboratory, NASA Goddard Space Flight Center, 8800 Greenbelt Rd., Greenbelt, MD 20771, USA}
\author{John G. Baker}
\affiliation{Gravitational Astrophysics Laboratory, NASA Goddard Space Flight Center, 8800 Greenbelt Rd., Greenbelt, MD 20771, USA}
\author{Bernard J. Kelly}
\affiliation{Gravitational Astrophysics Laboratory, NASA Goddard Space Flight Center, 8800 Greenbelt Rd., Greenbelt, MD 20771, USA}
\affiliation{CRESST and Department of Physics, University of Maryland, Baltimore County, 1000 Hilltop Circle, Baltimore, MD 21250, USA}
\date{\today}

\begin{abstract}

We investigate the precision with which
the parameters describing the characteristics and 
location of nonspinning black hole binaries
can be measured with the Laser Interferometer Space Antenna (LISA).
By using complete waveforms including the inspiral, merger and ringdown 
portions of the signals, we find that LISA will have far greater precision
than previous estimates for nonspinning mergers that ignored the
merger and ringdown.  Our analysis covers nonspinning waveforms with
moderate mass ratios, $\,q \geq 1/10$, and total masses $10^5\lesssim M/\MSun\lesssim 10^7$.  
We compare the parameter uncertainties using the Fisher matrix formalism,
and establish the significance of mass asymmetry
and higher-order content to the predicted parameter uncertainties 
resulting from inclusion of the merger.  
In real-time observations, the later parts of the signal lead to significant
improvements in sky-position precision in the last hours and even the final 
minutes of observation.
For comparable-mass systems with total mass  $M/\MSun\sim10^6$, 
we find that the increased precision 
resulting from including the merger is 
comparable to the increase in signal-to-noise ratio.
For the most precise systems under investigation,
half can be localized to within $\cal O$(10 arcmin), 
and 10\% can be localized to within $\cal O$(1 arcmin).  
\end{abstract}


\maketitle

\section{Introduction}

Gravitational waves carry a tremendous amount of information through the universe.  
It is the goal of the emerging field of gravitational wave astronomy to access that information 
and bring it to bear on the problems of astrophysics and cosmology. The current generation of 
gravitational wave detectors, such as the Laser Interferometer Gravitational-wave Observatory 
(LIGO) \cite{Abramovici:1992ah}, are focused on the detection of gravitational waves from isolated astrophysical systems. LIGO and its contemporaries will also provide some minimal information on the parameters of these systems such as their mass, luminosity distance, and approximate sky position. The quality of this information will be limited by the relatively low signal-to-noise ratios (SNR) expected 
for LIGO events \cite{Ajith:2009fz}.

The Laser Interferometer Space Antenna (LISA) \cite{LISA1}, a space-based detector of gravitational
waves in the milliHertz band, will detect the inspiral and merger of supermassive black hole
binaries (BHBs) with very large SNRs ($100\sim 10^{4}$)  out to redshifts of $z\sim10$ or greater 
\cite{Flanagan:1997sx, Baker:2006kr}. These large SNRs make it possible to extract a large amount 
of information from each event including mass, mass ratio, spins, orientation, luminosity distance,
and sky position. Because sources of gravitational waves are strongly dominated
by gravitational dynamics, and because the waves are expected to propagate through intervening matter
with little interaction, these observations may provide an unusually clean and direct measurement
of the source system parameters. Of particular interest are the distance and sky position parameters,
which will drive LISA's ability to narrow the set of candidate source galaxies or clusters for merger
events, potentially opening up a range of multi-messenger astronomy opportunities.  For instance, the coincident measurement
of gravitational and electromagnetic signatures from a single source galaxy (\ie standard sirens \cite{Holz:2005df})
will allow a direct measurement
of the redshift-luminosity relationship, thereby constraining the dark-energy equation of state.
While cosmological models predict that the dark-energy-dominated era began fairly recently at $z\sim1$, measurements
for much larger redshifts are not currently possible by other methods, and the standard siren method is limited
only by the achievable range of coincident electromagnetic observation.

Extracting this information requires a waveform model that can provide templates with sufficient 
fidelity to distinguish between signals with different parameters in the presence of instrumental 
noise. For BHBs, the complete waveform signal is traditionally divided into three different 
regimes: the inspiral, which can be described using post-Newtonian (PN) orbital dynamics; the 
ring-down, which can be treated using black-hole perturbation theory; and the merger, which
bridges the two and can be predicted using numerical relativity.

Ideally, an estimate of LISA's ability to measure the parameters of observed BHBs would include 
the information contained in the complete waveform.  
However, the difficulty associated with modeling 
the merger has led the majority of such studies (the exceptions being \cite{McWilliams_PhD}, \cite{Babak:2008bu}, 
and \cite{Thorpe:2008wh}) to include only the inspiral portion of the 
waveform.
Until recent advances in numerical relativity \cite{Pretorius:2005gq,Baker:2005vv,Campanelli:2005dd} 
opened the door to a complete understanding of
General Relativity's predictions for these signals, it was not clear whether theoretical knowledge 
about the final strongest moments of the signal would be available for system parameter measurements. 
A naive guess at the consequences of omitting the merger would be that the loss in 
parameter precision would be proportional to the loss in SNR.  This assumes that the two portions 
of the signal have equal density of information per unit SNR and that that information is 
independent.  

There are two reasons that are sometimes cited for expecting that the effect of the merger on 
parameter precision will be less than that on SNR. The first is that the merger encompasses very 
few GW cycles compared with the observed portion of the inspiral and it is expected that 
information content correlates with the number of cycles.  The second is that, in the 
low-frequency limit, the sensitivity of LISA to parameters such as sky position is entirely 
generated by the orbital motion of the LISA constellation and the merger is too short in duration
to experience a significant orbital modulation. 

In this work we investigate the significance of the merger to LISA parameter estimation using 
quasi-analytic waveforms that are tuned to match the results of numerical relativity.  We restrict 
ourselves to non-spinning binaries with moderate mass ratios ($q\leq1/10$) and explore the 
parameter space around a candidate system with total redshifted mass of $1.33\times 10^6 \MSun$, 
mass ratio $q=1/2$, and redshift $z=1$.  Astrophysically, we expect black holes to have spin.
While including spin 
\cite{Vecchio:2003tn, Lang:2006bz} and mergers each separately 
improve parameter estimation, it is not known how these effects
will combine.  Therefore, including both
spin effects and mergers will be an important followup to this investigation.

In section \ref{sec:meth}, we discuss the methods employed for generating models of the 
complete waveform signals
and the instrument, 
and for estimating parameter measurement precision. 
In section \ref{sec:res}, we examine LISA's ability to measure binary black-hole system
parameters.  The primary novelty of our results is that we assume theoretical knowledge
of the complete signal is applied in the observational analysis.  We examine the  
impact of including the merger and higher harmonic content (\ref{subsec:harm})
for comparable mass systems near $10^6 \MSun$, 
the variation of the results across a range of masses (\ref{subsec:mtot}) and mass ratios (\ref{subsec:mrat}).
In  (\ref{subsec:timeEvo}) we study how sky position information accumulates in time,
which will impact how LISA ultimately interacts with other astronomical instruments.
We summarize our key results in section \ref{sec:conc}.

\section{Methodology}
\label{sec:meth}

Before presenting our findings, we will briefly review the steps taken to estimate the 
precision with which LISA will measure astrophysical parameters using complete waveforms. 
Theoretical predictions of the strong-field gravitational dynamics and radiation generation
must be encoded in a parameterized waveform signal model.
We then need to apply a model of the instrument, including the
response to signals and the sources of noise.  Finally, we need to estimate the theoretical
limit on the uncertainty of the measured signal parameters that could be achieved
from a measurement consisting of our realizations of the signal and noise content.

\subsection{Waveform model}
\label{subsec:wmod}

We assume that Einstein's theory of gravity correctly describes black hole binary systems.
While numerical relativity can now treat the final moments of these events, it
would be impractical to conduct simulations covering the parameter space of interest.
Furthermore, general signal templates must cover the complete signal including the long-lasting
inspiral signal which can not be modeled numerically, but which is well-described by
the post-Newtonian (PN) approximation \cite{Blanchet06}.
Instead our complete waveform model is based on a variation of the PN treatment
that has been ``tuned'' to approximate the numerical simulation results at late times.
Such a model can be tuned using available numerical data, while providing reasonable, if unverified, 
model signals for arbitrary parameter sets.  

In order to investigate LISA's capabilities for recovering source parameters,
we specifically use a waveform model \cite{Baker:2008mj} tuned to match the available
numerical simulations for nonspinning black hole binaries.  This model, referred to as the 
IRS-EOB model, uses a conventional effective-one-body (EOB) Hamiltonian formalism for the
adiabatic inspiral \cite{Buonanno:2007pf}.  For the merger-ringdown, a fit to a physically-motivated
functional form is employed for the phasing (see Eq.~9 
in \cite{Baker:2008mj}), while the amplitude is calculated using a model for the flux
that is constrained both to be consistent with the inspiral flux through
3.5 PN order, and also to vanish as it approaches the ringdown frequency (referred
to as ``Model 2'' and given by Eq.~19 in \cite{Baker:2008mj}).  The physical motivation, that the radiation
can be treated as though it were being generated by a shrinking rigid rotator, explains the
IRS in IRS-EOB, which stands for ``implicit rotating source''.

For a unit mass-system ($m_1+m_2=1$), the source model depends only on the remaining intrinsic source
parameters, the mass ratio $q\equiv m_1/m_2$ (where $m_2>m_1$), and the
spins, which we set to vanish for this initial investigation.
Throughout this work, we employ waveforms that correspond to $\sim 10^6\, M$ of observation,
or $\sim 3$ months for our 
fiducial case with $M=1.33\times 10^6\,\MSun$.  Here, we use units of $G=c=1$, so that 
$1\,M = 4.92\times10^{-6}\,(M/\MSun)$ seconds.  Therefore,
lower-mass systems require longer simulations in $M$.
We are limited computationally from employing longer waveforms, but
we have verified that our results do not change significantly (with the single
exception of the uncertainty in total system mass) by doubling the waveform
length for the lowest-mass cases investigated.  We use a model cadence
of $0.5\, M$.  The signal is resampled when we apply the detector's response function,
so that the final signal cadence corresponds to a quarter wavelength
at the highest frequency reached by the $\ell=4$, $m=\pm4$ harmonics.  This
is true even for cases where we restrict the calculation to have only quadrupolar
content.

After the source calculation, we derive the incident waveforms 
referenced to the solar system barycenter (SSB),
at which point we can apply the response of the LISA detector.
Computation of the incident waveform in SSB frame depends on eight additional
parameters: the redshifted total system mass $M=M_o(1+z)$ 
(with $M_o$ the rest mass and $z$ the redshift), 
luminosity distance $D_{L}$, coalescence time $t_{c}$, and three angles describing
the orientation of the binary, for which we use the inclination $\iota$ 
(using the convention that $\iota=0$ corresponds to the line of sight 
being coincident with the orbital axis of the binary),
initial orbital phase $\phi_{o}$ and the polarization phase $\psi$
At the source, the emitted radiation can be decomposed in spin-weighted
spherical harmonic components $h_{\ell m}$
of the dimensionless gravitational 
wave strain (scaled for unit distance from the source). 
Here the strain is complex-valued to represent both polarization components, 
$h\equiv h_{+}\,+\,i h_{\times}$.
Specifying the parameters $(\iota,\phi_0,\psi,M, D_L)$ 
allows us to calculate the solar-system incident waveform $\hB$:
\beq
\hB = \frac{GM}{c^2D_L}\left[e^{2i\psi}\sum_{\ell m}\, ^{-2}Y_{\ell m}(\iota,\phi_o)\,h_{\ell m}\left(\frac{t_c-t}{M}\right)\right]\,,
\label{eq:source2SSB}
\eeq
where $^{-2}Y_{\ell m}$ are the spin-weight $-2$ spherical harmonics \cite{Goldberg:1967}.
The two additional parameters, the ecliptic latitude $\beta$ and longitude $\lambda$,
describe the sky location of the binary in the SSB frame.
The dependence on sky location is applied by the instrument response, which we discuss below.
We use the vector
$\Lambda^a\equiv(\ln M,\ln D_{L},\beta,\lambda,\iota,\phi_{o},\psi,t_{c})$ to
denote the complete set of variable parameters.  Note that the dependence on mass ratio, $q$, is not explicitly included in $\Lambda^a$ but is instead implicitly included in the $h_{lm}$. This is because $q$ is not varied when computing parameter uncertainties (see section \ref{subsec:fish}), a procedure consistent with that used in \cite{Babak:2008bu}. This is equivalent to the assumption that there is no 
uncertainty in the measurement of $q$. We intend to relax this assumption in future investigations. 

\subsection{Instrument model}
\label{subsec:imod}

The instrument model consists of two components: a prescription for converting $\hB$ 
into signals observed by the instrument, and a description of the instrument noise.  The LISA instrument consists of a constellation of three spacecraft 
located at the vertices of an approximately equilateral triangle with a side length of $5\times 10^{9}\,\mbox{m}$.  The light travel time along each 
of the six one-way links is monitored using laser interferometry. These individual link measurements are then combined using a technique known as Time 
Delay Interferometry (TDI) \cite{Tinto:1999yr} to yield observables that contain gravitational wave signals and suppress instrumental noise. 
Of the many families of TDI observables \cite{Tinto:2003uk}, the ones most suitable for data analysis are the orthogonalized or ``optimal'' variables \cite{Prince:2002hp}. 
For this investigation, we have developed a set of orthogonal variables we refer to as ``pseudo-{$A$, $E$, $T$}'' (hereafter $\bA\equiv\{\bar{A},\bar{E},\Tbar\}$), 
which are analogous to the original ${\A} \equiv\{A,E,T\}$ variables in \cite{Prince:2002hp}  
except that they are constructed from the Michelson  ${\X} \equiv \{X,Y,Z\}$ variables 
rather than the TDI generators $\balpha \equiv\{\alpha,\beta,\gamma\}$.  The software package \textit{Synthetic LISA} \cite{Vallisneri:2004bn} is used to generate the 
$\X$ variables from the incident gravitational waveforms, $\hB$ and the $\bA$ variables are then computed as
\bea
\bar{A}&=& \frac{Z-X}{2\sqrt{2}} \nonumber \\
\bar{E}&=& \frac{X+Z-2\,Y}{2\sqrt{6}} \nonumber \\
\Tbar&=& \frac{X+Y+Z}{2\sqrt{3}}\,.
\eea
An overall factor of $\frac{1}{2}$ has been applied to all three formulas to make $\bar{A}$ and $\bar{E}$
agree with $A$ and $E$ in the low-frequency limit. 

\textit{Synthetic LISA} can also be used to model instrument noise.  However, 
\textit{Synthetic LISA} includes statistical fluctuations in its noise-generation algorithms,
whereas we are interested in studying parameter uncertainties
that result from differences in the incident waveforms for typical instrumental noise levels.
We therefore wish to suppress the impact of statistical fluctuations
in a given realization of the noise. One way to do this is to generate an ensemble of noise realizations and average them.  Satisfactory results can be achieved with a suitable number of averages at the expense of increased computational effort.  For this analysis, we have followed the procedure used in \cite{Estabrook:2000ef} to produce analytic estimates of the mean power spectral densities of the noise in the TDI channels directly from the acceleration and optical path length noises in the individual 
links. This procedure requires making assumptions such as stationary, equal arm lengths. The expressions for the one-sided spectral densities in the $\bA$ observables are
\begin{widetext}
\bea
S_{\bar A, \bar E}&=&2\sin^2(\Phi)\left[2\left(3+2\cos(\Phi)+\cos(2\,\Phi)\right)S_{\rm pm}+\left(2+\cos(\Phi)\right)S_{\rm op}\right], \nonumber \\
S_{\Tbar}&=&8\sin^2(\Phi)\sin^2(\Phi/2)\left[4\sin^2(\Phi/2)S_{\rm pm} + S_{\rm op}\right]\,,
\label{eq:Saet}
\eea
\end{widetext}
where $\Phi \equiv \omega L/c$ and
$L$ is the arm length, expressed as a light-travel time. 
These expressions are the analog of Eqs.~67 and 68 in \cite{Krolak:2004xp} and Eqs.~19 and 20 in \cite{Prince:2002hp}
for the noise response of the original $\A$, and we have verified that we 
duplicate the results in \cite{Prince:2002hp} for $\A$ using $\balpha$ (accounting for a typographical error
which appears in Eq.~20 of \cite{Prince:2002hp} which, if corrected, would make it consistent with
\cite{Krolak:2004xp} and with our results).

The quantities $S_{\rm pm}$ and $S_{\rm op}$ are the one-sided
spectral densities of the proof mass acceleration and optical path-length noises, 
respectively, expressed as equivalent strain.
They are modeled as
\bea
S_{\rm pm} &=& 2.5\times 10^{-48}\left( \frac{f}{1\,\mbox{Hz}}\right)^{-2}\sqrt{1+\left(\frac{f}{0.1\,\mbox{mHz}}\right)^{-2}}, \nonumber \\
S_{\rm op} &=& 1.8\times 10^{-37}\left( \frac{f}{1\,\mbox{Hz}}\right)^{2}.
\label{eq:Spmop}
\eea
The acceleration noise power spectrum in $S_{\rm pm}$ includes an additional $f^{-1}$ reddening 
below $0.1$ mHz to account for the unmodeled behavior of the instrument below the LISA band.

\begin{figure}
\includegraphics[trim = 0mm 0mm 0mm 0mm, clip, scale=.16, angle=0]{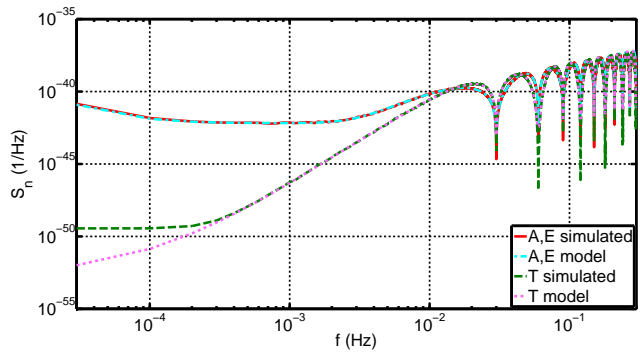}
\caption
{Comparison of different noise realizations for the {$\bar{A}$, $\bar{E}$, $\Tbar$} channels.  
We have employed a noise model (light dash-dotted for $\bar{A}$ and $\bar{E}$, 
light dotted for $\Tbar$), 
and verified that it agrees well with
an averaged ensemble of simulated noise using \textit{Synthetic LISA} (dark solid for $\bar{A}$ 
and $\bar{E}$, dark dashed for $\Tbar$). Galactic foreground noise is not included in these traces
but is included in parameter sensitivity calculations. \vspace{10mm}}  
\label{fig:noise}
\end{figure}

As a check of the expressions in Eq.~\ref{eq:Saet}, we used \textit{Synthetic LISA} to model 
an ensemble of 1000 realizations of noise in the TDI $\bA$ channels. Fig.~\ref{fig:noise} shows a 
comparison of the mean power spectra of this ensemble with the expressions in Eq.~\ref{eq:Saet}.  
In general, the agreement between the simulated noise and the analytic expressions is quite 
good. Deviations between the two curves indicate areas of potential concern when evaluating 
SNR and parameter sensitivity.  For example, the analytic noise expressions in Eq.~\ref{eq:Saet} 
contain nulls at frequencies corresponding to the inverse round-trip times of the constellation. 
The simulated data, be it signal or noise, is finite at these frequencies due to 
spectral estimation effects. This 
leads to a spurious divergent contribution to the SNR and parameter uncertainties
at these frequencies.
To guard against this, we applied a noise floor of $10^{-40}(f/1\,\mbox{Hz})^{2}$, eliminating the nulls in the noise response.  In addition the `flexing' of the LISA arms due
to orbital variations was disabled in \textit{Synthetic LISA} to maintain consistency with the expressions in
Eq.~\ref{eq:Saet}, which were derived assuming constant arm lengths.

\vspace{10mm}

The other area of disagreement between the simulated and analytic noise in 
Fig.~\ref{fig:noise} is at the low-frequency end of the $\Tbar$ channel.  The analytic 
expression in Eq.~\ref{eq:Saet} predicts that the noise in the 
$\Tbar$ channel should continue to decrease 
with decreasing frequency, while the simulated noise levels off. 
As a result of using \textit{Synthetic LISA} to model the signal response, and
Eq.~\ref{eq:Saet} to model the noise, there would be a
large spurious contribution to SNR and parameter sensitivity at low frequencies, precisely
the band where the $\Tbar$ 
channel is not expected to contribute.  As the source of this discrepancy has not yet been 
identified, we have elected to exclude the $\Tbar$ channel in the remainder of our analysis.
We note that, since the information from $\Tbar$ will be present at high frequencies, its exclusion
will lead us to produce conservative uncertainty estimates with systematically worse 
uncertainties than might be otherwise obtained for higher-mass 
cases where $\Tbar$-channel response can be non-negligible.
Our treatment is consistent with previous studies which have generally neglected the details of
LISA high-frequency response. We plan to include $\Tbar$ in future work.

The final component of the noise model is the foreground of gravitational waves from 
unresolved compact binaries.  We use the model for the galactic foreground that was 
developed in \cite{Timpano:2005gm}. Specifically we add to the expressions for $S_{\bar A, \bar E}$ 
in Eq.~\ref{eq:Saet} a galactic foreground noise, $S_{\rm gal}$, given by
\beq
S_{\rm gal} = [4\Phi\sin(\Phi)]^2 \,S_{\rm conf}\,,
\label{eq:Sgal}
\eeq
where $S_{\rm conf}$ is taken from Eq.~14 of \cite{Timpano:2005gm}. The contribution from 
$S_{\rm gal}$ is not included in Fig.~\ref{fig:noise} but is included in all SNR and 
parameter estimation calculations.

\subsection{Parameter estimation using the Fisher matrix}
\label{subsec:fish}

To approximate the measurement precision that LISA can achieve, one approach we can 
take is the Fisher matrix formalism.  If the LISA data stream consists of a waveform, 
$h(\Lambda^a)$, embedded in a signal, $s$,
so that the noise, $n$, is given by $n = s - h$, then the probability that a signal 
contains a waveform with the parameter set $\tilde{\Lambda}^a$ is given by the likelihood function,
\beq
p(\tilde{\Lambda}^a|s) \propto e^{-\langle h(\tilde{\Lambda}^a) - s|h(\tilde{\Lambda}^a) - s\rangle/2}
\label{eq:prob}
\eeq
where $\langle \cdots | \cdots \rangle$ is a noise-weighted inner product \cite{Cutler:1994ys}.
The ``maximum likelihood'' set of parameters, $\hat{\Lambda}^a$, 
is the one that maximizes $p$.  Errors in the $\hat{\Lambda}^a$ set
of parameters can be assessed by expanding $p$ around $\hat{\Lambda}^a$, such that
\beq
p(\tilde{\Lambda}^a|s) \propto e^{- \Gamma_{ab}\delta\Lambda^a\delta\Lambda^b/2}
\label{eq:probfish}
\eeq
where $\delta\Lambda^a \equiv \tilde{\Lambda}^a - \hat{\Lambda}^a$.  The Fisher information matrix, 
$\Gamma_{ab}$, which is
the centerpiece of our subsequent analysis, is defined to be
\beq
\Gamma_{ab} \equiv \bigg\langle \frac{\partial h}{\partial \Lambda^a}\, \bigg| \,
\frac{\partial h}{\partial \Lambda^b }\bigg\rangle,
\label{eq:fish}
\eeq
where $a$ and $b$ are parameter indices.  Throughout this work, we calculate the
parameter derivatives in Eq.~\ref{eq:fish} using one-sided differencing,
with a fractional step size $\varepsilon^a=\Delta\Lambda^a$, where we set $\Delta=10^{-4}$
for the coalescence time, and $\Delta=10^{-6}$ for all other parameters.

To lowest order in an expansion in SNR$^{-1}$, the covariance matrix, $\Sigma^{ab}$,
is just the inverse of the Fisher matrix:
\beq
\Sigma^{ab} = \left(\Gamma^{ab}\right)^{-1} \left[1 + {\cal O}({\rm SNR}^{-1})\right],
\label{eq:cov}
\eeq
so that $\sigma^a \equiv \sqrt{\Sigma^{aa}}$ is the standard deviation of parameter $a$.
The covariance matrix is symmetric, with
the off-diagonal terms giving the covariance between parameters, and the diagonal terms giving 
the variance of each parameter.  Because inverting the Fisher matrix to find the covariance
matrix is not always valid, we verify our results by testing individual cases at random for
each system of interest using Markov Chain Monte Carlo simulations.

Because we use $\ln M$ and $\ln D_L$ as parameters,
the resulting uncertainties are fractional:
\bea
\sigma^{\ln M} &\approx& \sigma^M/M\,, \nonumber \\
\sigma^{\ln D_L} &\approx& \sigma^{D_L}/D_L\,.
\label{eq:lnsig}
\eea
We therefore express these as fractional uncertainties throughout this work.
Though we will refer to the quantities $\sigma^a$ as ``uncertainties'' throughout this work,
we wish to note that they are a measure of precision, not necessarily accuracy.
 
Another parameter of interest is the precision of the sky localization, expressed as the
area of the uncertainty ellipse on the sky, $\Omega$.  This is sometimes referred to as
$\Delta\Omega$ or $\sigma^{\Omega}$ in the literature, 
but we simply use $\Omega$, given that it is a measure
of an area of uncertainty, rather than a measure of uncertainty of an area.
To construct this value from
our data, we make the approximation
\bea
\label{eq:omega}
\Omega &=& 2\pi\sqrt{\left(\sigma^{\lambda}\,\sigma^{\cos\beta}\right)^2 - \left(\Sigma^{\lambda,\,\cos \beta}\right)^2} \\ \nonumber
&\approx& 2\pi\sin\beta\sqrt{\left(\sigma^{\lambda}\,\sigma^{\beta}\right)^2 - \left(\Sigma^{\lambda,\,\beta}\right)^2}\,.
\eea
We note that generally $\Omega \neq \pi\,\sigma^{\lambda}\,\sigma^{\beta}$, both because
we quote median values throughout, and because we define the ellipse as in \cite{Cutler:1997ta},
so that the effect of Eq.~\ref{eq:omega} is to diagonalize the Fisher sub-matrix
formed by $\beta$ and $\lambda$, and thereby calculate the area using the true semi-major
and semi-minor axes.

\subsection{Information Accumulation}
\label{subsec:infoAccum}

\begin{figure}
\includegraphics[trim = 0mm 0mm 0mm 0mm, clip, scale=.16, angle=0]{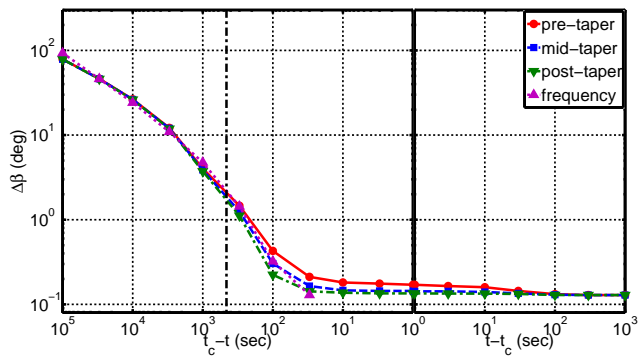}
\caption
{Estimated improvement in ecliptic latitude uncertainty as a function of signal truncation 
time, $t$, relative to the coalescence time, $t_c$ for an equal-mass merger with 
$M=1.33\times 10^6 \,\MSun$ at $z=1$. 
The curves labeled ``pre-taper'', ``mid-taper'', and 
``post-taper'' were computed using time-domain truncation, with the
truncation time corresponding to the end of, middle of, and beginning of
a one-radiation-cycle-long taper, respectively.  The 
curve labeled ``frequency'' was computed using the full waveform but imposing an upper 
frequency cutoff when evaluating the inner product in Eq.~\ref{eq:fish} corresponding to 
the instantaneous signal frequencies at the truncation times. The vertical dash-dotted 
line corresponds to the Schwarzschild ISCO (see Sec.~\ref{subsec:harm}).  
The vertical thick solid line separates
the times prior to merger from the times after merger.
\vspace{10mm}}
\label{fig:timefreq}
\end{figure}

One aspect of LISA's parameter sensitivity that is of interest is the way in which 
parameter uncertainty evolves with time. A simple way to investigate this is to truncate
the signal at a specified sequence of times before merger.  To avoid edge 
effects in the ensuing spectral estimation, the truncated signal is tapered with a raised 
cosine window with a length of approximately one wave cycle of the quadrupole mode at
the time of truncation. The 
Fisher matrix and covariance matrix are then computed using this truncated signal.  
By constructing a sequence with progressively later truncation times,
one can trace the evolution of 
parameter uncertainty.  Fig.~\ref{fig:timefreq} shows the time-evolution of the uncertainty
in the ecliptic latitude, $\beta$,
for an equal-mass system with $M=1.33\times 10^6 \MSun$ 
at $z=1$. Three variants of the time curves are shown corresponding to different 
tapering conventions. The taper either starts at the designated truncation time,
ends at that time, or the mid-point of the taper occurs at that time.  This distinction,
while seemingly trivial, shows the potential impact of slight differences in the treatment of
time-domain signals with regard to taper length, type, or placement, 
as clear differences can be seen in Fig.~\ref{fig:timefreq}.
For early times, the taper 
prescription does not matter.  At later times, the precise taper implementation 
becomes more important, leading to a time shift between the three curves. In our discussion of 
information accumulation in section \ref{subsec:timeEvo}, truncation time refers to the
mid-point of the taper.

For signals comprising a single harmonic (or rather, a single pair of $\ell,\,\pm m$ 
modes, which are complex conjugates of each other for nonspinning waveforms), 
it is straightforward to repeat this analysis in the frequency domain. This provides an
internal consistency check for our code. For each desired truncation time, the 
instantaneous frequency
of the source waveform is used to compute the corresponding 
signal frequency. The evaluation of the inner product in Eq.~\ref{eq:fish} is 
then limited to frequencies below this.  The curve in Fig.~\ref{fig:timefreq} shows that 
this approach is consistent with the time-domain approach, at least up to times 
very near the merger. It would be possible to extend this technique by tracking 
each mode separately and computing a different frequency cutoff for each. However, 
we find the time-domain approach to be more straightforward.

\vspace{10mm}

\subsection{Caveats}
\label{subsec:codeval}

It is well known that the Fisher-matrix approach is prone to a number of potential
pitfalls \cite{Vallisneri:2007ev}. In this section, we attempt to address a few of them.  

The first potential issue is our approximation of the parameter derivatives,  
$\partial h/\partial \Lambda^a$, using a one-sided finite difference approach. 
As a rough check of the validity of this approximation, we have computed the Fisher 
and covariance matrices using various finite difference step sizes and verified 
that the results were consistent. Table \ref{Table:covconv} demonstrates that an 
order of magnitude decrease in the step size changes covariance terms by a few percent at worst.

\begin{table}
\begin{center}
\begin{tabular}{c c c c c c c c}
\hline \hline
$M$ & $D_L$ & $\beta$ & $\lambda$ & $\iota$ & $\phi_o$ & $\psi$ & $t_c$\\
4.8e-2 & 5.7e-2 & 1.3e-2 & 5.0e-2 & 4.9e-2 & 4.8e-2 & 6.8e-4 & 4.4e-2\\
 - & 3.0e-5 & 1.9e-4 & 2.6e-5 & 1.0e-3 & 9.1e-4 & 1.9e-4 & 4.0e-5\\
 - & - & 1.4e-3 & 1.3e-3 & 4.0e-4 & 2.1e-4 & 3.9e-4 & 1.5e-4\\
 - & - & - & 1.3e-4 & 5.2e-5 & 7.1e-3 & 1.5e-4 & 5.5e-5\\ 
 - & - & - & - & 5.1e-5 & 1.1e-2 & 1.3e-3 & 1.3e-4\\ 
 - & - & - & - & - & 1.6e-4 & 2.1e-4 & 7.1e-3\\ 
 - & - & - & - & - & - & 4.0e-4 & 5.2e-5\\ 
 - & - & - & - & - & - & - & 3.4e-4\\
\hline \hline
\end{tabular}
\end{center}
\caption
{Consistency of the covariance matrix for an equal-mass system with 
$M=1.33\times 10^6\,\MSun$ at $z=1$. The 
tabulated quantity is the symmetric matrix $|1-\Sigma_{\varepsilon_1}/\Sigma_{\varepsilon_2}|$, 
where $\Sigma_{\varepsilon_i}$ is a covariance matrix calculated using the Fisher 
matrix method with a one-sided finite-difference step size of $\varepsilon_i$. For this
comparison, $\varepsilon_1 = 10^{-6}$ and 
$\varepsilon_2 = 10^{-7}$. \vspace{10mm}}
\label{Table:covconv}
\end{table}

A related concern is that the Fisher information matrix precision estimates
assume that the relevant portion of the likelihood function can be treated as
a quadratic function. This assumption should be guaranteed by the large SNR, but is not 
explicitly verified.

Finally, actual observations will require the implementation of concrete algorithms
for exploring the likelihood function, such as those pursued in the Mock LISA
Data Challenges \cite{MLDC}.  In those Challenges, it has been demonstrated
that accuracy may be impacted, for instance, by complicated structure 
in the likelihood function, including multiple maxima and extended shallow 
regions.  Systematic errors are also possible, for instance, 
if errors in the theoretical signal predictions should exceed
statistical errors \cite{Cutler:2007mi}.  
Highly accurate merger waveform predictions, and corresponding models tuned to those
predictions, are currently
available only for a very limited sampling of specific black hole system 
configurations.  However, this area of study is advancing rapidly, and 
it now appears that accurate information
about the complete signals throughout the relevant parameter space is likely to be available
at the time of LISA's operation.  We therefore focus on the additional source information
that may be obtainable when the full signal predictions are applied 
in the observational analysis.
 
\section{Results}
\label{sec:res}

Since there are variations in the parameter uncertainties across the 
parameter space, we perform
Monte Carlo simulations to find the distribution of uncertainties
For specific choices of masses and luminosity distance 
(parameters $M$, $q$, and $D_L$), 
we conducted Monte Carlo simulations consisting of $1024$ randomly-generated
parameter sets for all the cases shown, and have spot-checked that our 
results do not change significantly if we increase to $8192$ parameter sets.  
The remaining parameters are drawn from uniform 
distributions, with $\iota$ drawn from a uniform distribution
in $\cos\iota$ to give uniform sky coverage.

\begin{table*}
\begin{center}
\begin{tabular}{c c c c|c c c c c c c c c c c}
\hline \hline
$(1+z)M$ & $m_1/m_2$ & Numerator & Denominator & $\sigma^M/M$ & $\sigma^{D_L}/D_L$ & $\sigma^{\beta}$ & $\sigma^{\lambda}$ & $\Omega$ & $\sigma^{\iota}$ & $\sigma^{\phi_o}$ & $\sigma^{\psi}$ & $\sigma^{t_c}$ & $\mathrm{SNR}^{-1}$ \\
\hline
 1.33e6 & 1/1  & $\ell\leq\,4$, ISCO & $\ell\leq\,4$, full & 1.2 & 3.6 & 4.8  & 6.4 & 27 & 2.3 & 2.7 & 5.0 & 13 & 3.1 \\
 - & -  & $\ell=|m|=2$, full & $\ell\leq\,4$, full & 1.0 & 2.3 & 1.8 & 1.5 & 2.7 & 1.8 & 1.7 & 2.5 & 1.2 & 1.0 \\
 - & -  & $\ell=|m|=2$, ISCO & $\ell\leq\,4$, ISCO & 1.0 & 9.1 & 6.1 & 4.0 & 28 & 8.3 & 6.8 & 7.1 & 1.2 & 1.0 \\
\hline
 1.33e6 & 1/2 & $\ell\leq\,4$,ISCO  & $\ell\leq\,4$, full & 1.2 & 3.5 & 5.5 & 5.6 & 29  & 2.9 & 2.7 & 4.8 & 15  & 3.2 \\
 - & -        & $\ell=|m|=2$, full  & $\ell\leq\,4$, full & 1.0 & 4.5 & 3.1 & 2.3 & 8.2 & 3.8 & 3.9 & 4.8 & 1.3 & 1.0 \\
 - & -        & $\ell=|m|=2$, ISCO  & $\ell\leq\,4$, ISCO & 1.0 & 16  & 8.1 & 6.1 & 63  & 16  & 17  & 14  & 1.3 & 1.0 \\
\hline \hline
\end{tabular}
\end{center}
\caption{Ratio of the
  waveform-model results for median variance of all the extrinsic
  parameters for two sets of comparable-mass physical systems.  The
  ``Numerator'' and ``Denominator'' columns indicate the models
  compared in constructing the ratios for that row.  The models vary
  by the harmonic content of the waveforms and by whether the merger
  is included (full) or not (ISCO).  For the systems considered here,
  the fractional loss in estimated precision from ignoring the final merger is
  comparable to the corresponding loss in SNR and has a greater impact
  than ignoring higher harmonics.  The significance of the higher
  harmonics is lower when full models are considered, as compared with
  ISCO-terminated models. The actual median fractional variances for
  all cases are given in Table \ref{Table:varcomp}.
\vspace{10mm}}
\label{Table:ratio}
\end{table*}

\subsection{Adding the merger and higher harmonics}
\label{subsec:harm}

Table \ref{Table:ratio} summarizes the improvement
in parameter uncertainties resulting from the addition of merger
for the cases of equal-mass systems and mass-ratio $q=1/2$, each
with a total mass   of $1.33\times10^6\MSun$ at a redshift $z=1$.  
We compare the uncertainty estimates obtained with four different
options for the waveform models.
Two of these options consist of the $\ell=2, m=\pm 2$
modes only, with one tapered in time to remove the merger, 
and the other including the full inspiral-merger-ringdown signal.
The midpoint of the taper corresponds to the time when the signal
reaches the frequency of the innermost stable circular orbit (ISCO) frequency of a test particle orbiting a Schwarzschild black hole, 
$f_{\rm ISCO} = c^3/\left(6^{3/2}\pi GM\right)$.  Much of the previous 
systematic work on
parameter uncertainties with LISA observations has applied waveform models
similar to the ($\ell=|m|=2$, ISCO) option.
More recent work has included higher harmonic
content\cite{Arun:2007hu,Trias:2008prd,Trias:2008pu,Arun:2008xf,Porter:2008kn}. Our other two waveform options include modes up to 
$\ell\leq 4$, where one case is again tapered to remove the merger,
and the other includes the complete signal.

The top row for each system included in Table \ref{Table:ratio} shows the
ratio of parameter uncertainties with and without the merger when higher
harmonics are included. In each case the inclusion of the merger
increases the SNR by roughly a factor of 3.  In general terms, if the 
information contained in the merger waveform is equally rich, as compared 
with the inspiral waveform, the uncertainties should decrease by a similar
factor.  This is generally the case for most parameters with the mass
uncertainty showing the least improvement and the sky-position, polarization
and coalescence-time uncertainties improving most.

The second and third rows for each system in Table \ref{Table:ratio}
summarize the improvement in uncertainties resulting from the inclusion 
of higher-harmonic content in the waveforms.  The second row shows the 
improvement in uncertainty when the full waveform is included in each model, 
while the third row shows the effect of including the higher harmonics 
with waveform models terminating at ISCO.  The latter comparison is 
roughly similar to previous considerations of the impact of higher 
harmonics \cite{Arun:2007hu,Trias:2008pu,Arun:2008xf}.
Comparing the second and third rows provides some indication of the
independence of information in the higher-harmonics and in the post-ISCO
merger. For most parameters the marginal effect of including higher harmonics
is not as great when the full-length waveforms are considered as it was for
ISCO-terminated waveform models.

\begin{figure*}
\begin{center}
\includegraphics*[trim = 0mm 0mm 0mm 8mm, clip, scale=.35, angle=0]{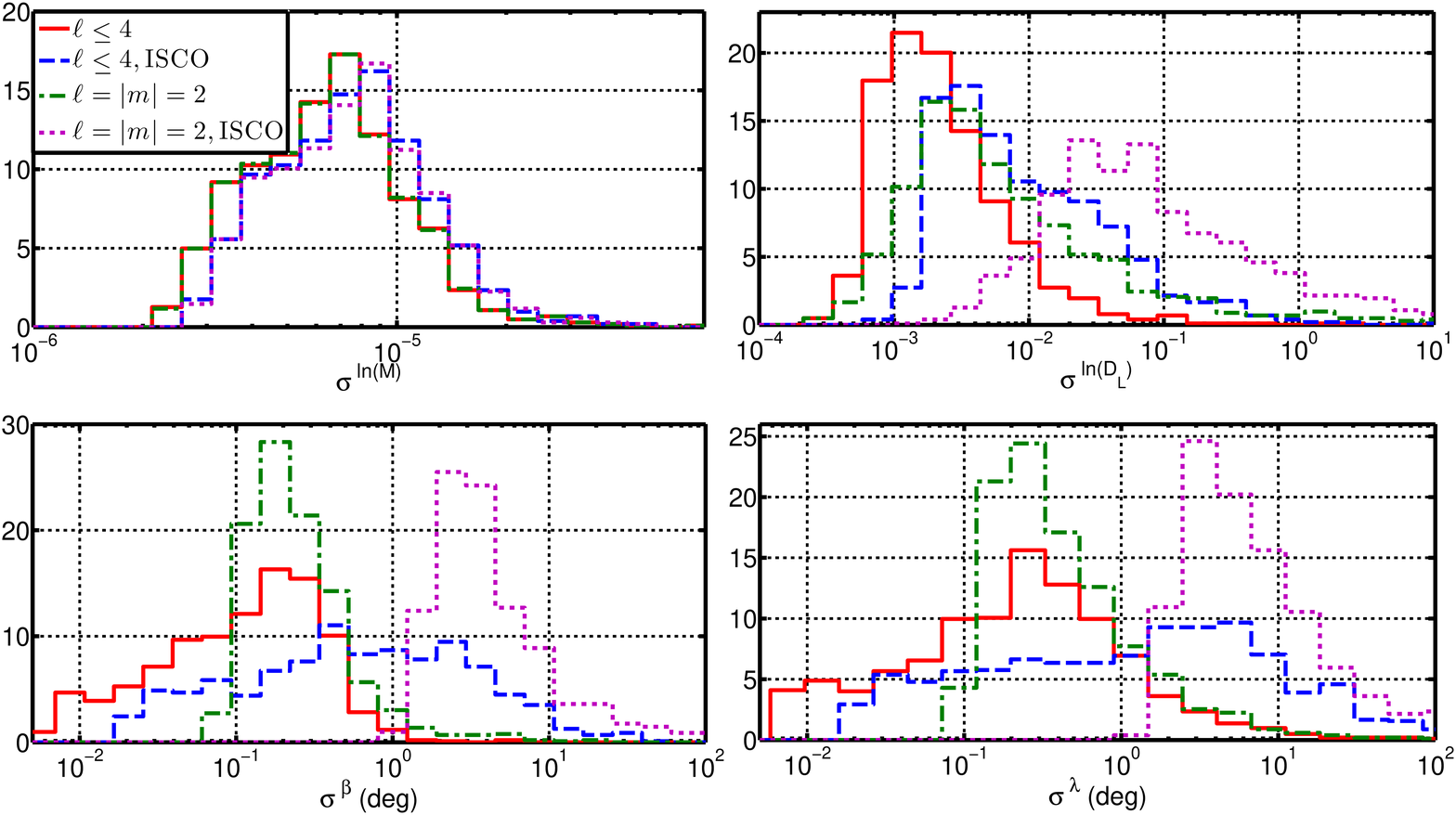}
\includegraphics*[trim = 0mm 0mm 0mm 8mm, clip, scale=.35, angle=0]{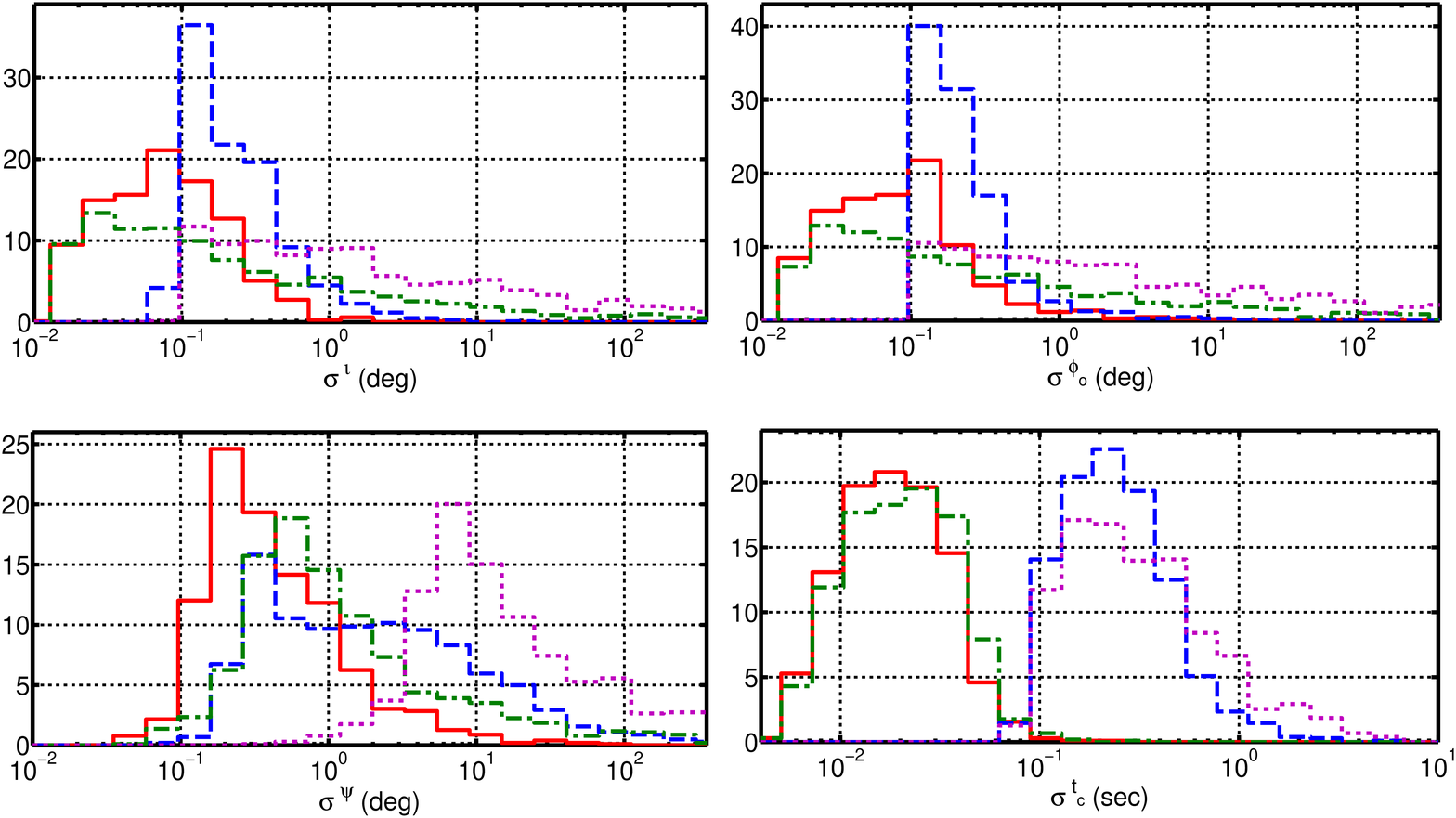}
\caption
{Uncertainty histograms for $q=1$ at redshift $z=1$, corresponding to a total system mass
$M=1.33\times10^6\MSun$, calculated using a
full inspiral-merger-ringdown waveform with harmonics $\ell\leq 4$ (solid),an inspiral waveform truncated at the ISCO as described in Sec.~\ref{subsec:harm} (dashed),
a full waveform including only quadrupole ($\ell=2$, $m=\pm 2$)
modes (dash-dotted), and an inspiral waveform
including only quadrupole modes (dotted).  All histogram bins are normalized by the total number of cases and are expres
sed as percentages.
\vspace{10mm}}
\label{fig:histsx1}
\end{center}
\end{figure*}

\begin{figure*}
\begin{center}
\includegraphics*[trim = 0mm 0mm 0mm 8mm, clip, scale=.35, angle=0]{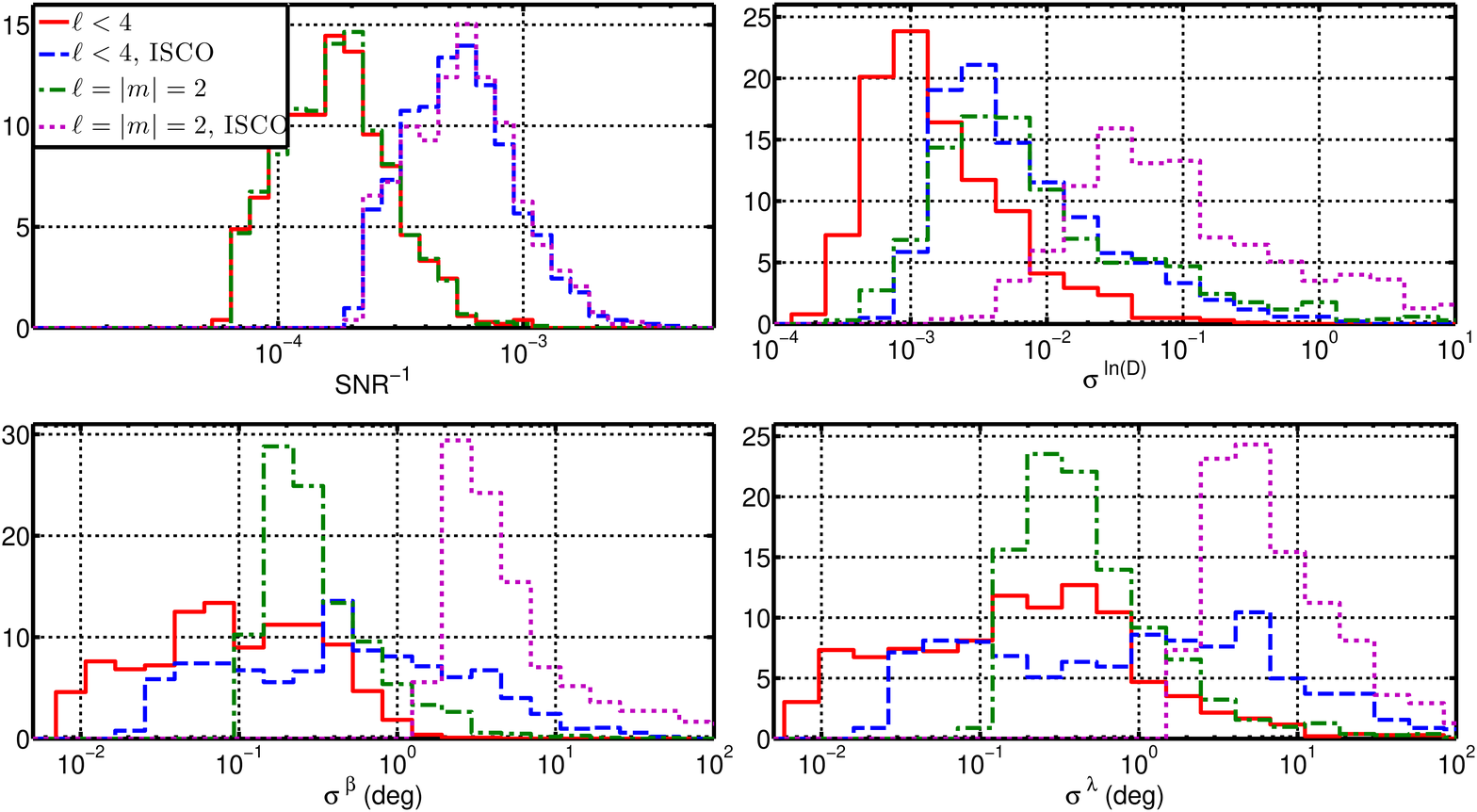}
\includegraphics*[trim = 0mm 0mm 0mm 8mm, clip, scale=.35, angle=0]{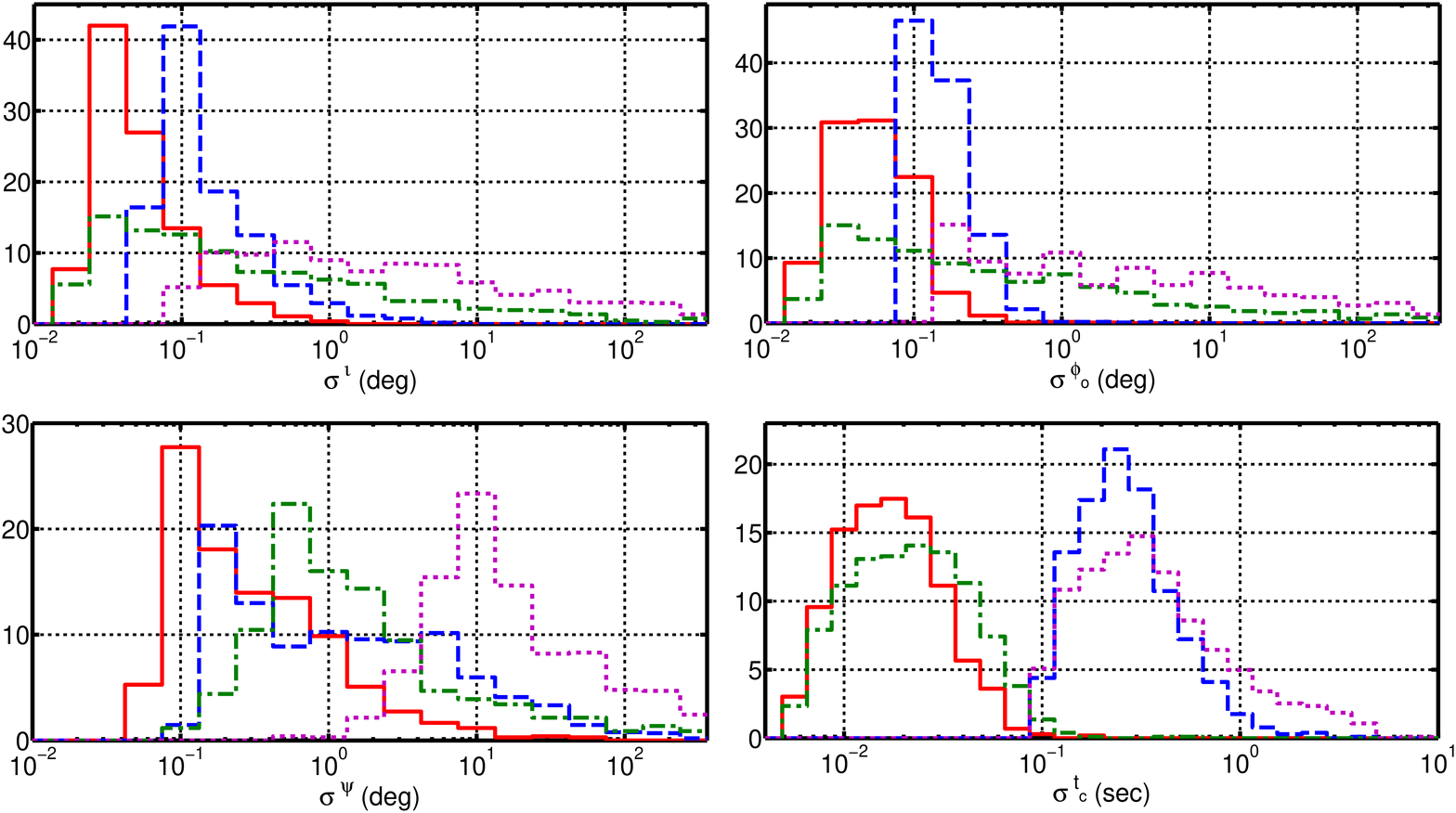}
\caption
{Uncertainty histograms for $q=1/2$ at redshift $z=1$,
corresponding to a total system mass $M=1.33\times10^6\MSun$,
calculated using a
full inspiral-merger-ringdown waveform with harmonics $\ell\leq 4$ (solid),
an inspiral waveform truncated at the ISCO as described in Sec.~\ref{subsec:harm} (dashed),
a full waveform including only quadrupole ($\ell=2$, $m=\pm 2$) modes (dash-dotted),
and an inspiral waveformincluding only quadrupole modes (dotted).
All histogram bins are normalized by the total number of cases and are expressed as percentages.
\vspace{10mm}}
\label{fig:histsx2}
\end{center}
\end{figure*}

In Figs.~\ref{fig:histsx1} and \ref{fig:histsx2}, 
we show histograms of our results for the four waveform model options for the
$q=1$ and $q=1/2$ systems, respectively.
The histograms in Fig.~\ref{fig:histsx1} agree qualitatively with those presented in our prior 
work \cite{Thorpe:2008wh}. Quantitatively there is some disagreement, which may be 
attributed to several 
factors. Chief among these are the increased duration of the signal ($\sim$10 days 
in the prior work 
as opposed to $\sim$3 months for a comparable  mass in this work) and an error in the 
prior code that omitted a factor of 
the TDI cadence in the parameter uncertainty estimates. 

For both the equal-mass case in Fig.~\ref{fig:histsx1}, and $q=1/2$ in Fig.~\ref{fig:histsx2}, 
we see a clear improvement in the level of measurement precision one can expect by including
the merger waveform.  It appears that, in particular, the parameter $t_c$ is localized extremely
well in both Figs.~\ref{fig:histsx1} and \ref{fig:histsx2} for cases that include the merger, 
relative to the timing precision without the merger.
Indeed, for every mass ratio the inclusion of
the merger is estimated to result in uncertainties in $t_c$ that are
an order of magnitude or more smaller than the smallest gravitational 
wave half-period reached by the signal waveforms,
which is the shortest time interval over which the signal will contain information content.
This dramatic improvement in timing accuracy can be heuristically explained by noting that the merger 
provides a sharp feature that can be well localized.  

The total system mass, $M$, is essentially insensitive to the inclusion of the merger
or the presence of higher harmonics, and appears to depend entirely on the
number of inspiral cycles, as was anticipated in \cite{Cutler:1994ys}. 
For this reason, 
we instead show $\mathrm{SNR}^{-1}$ in Fig.~\ref{fig:histsx2} and in subsequent histograms.  This quantity is useful, as it shows the degree of relative improvement
in parameter measurement that can be explained by an increase in SNR alone.

The precision of the luminosity distance $D_L$ and polarization phase $\psi$ measurements improve by 
roughly an order-of-magnitude over the quadrupolar inspiral case as either the merger or higher harmonics 
are added individually. When both features are added simultaneously, the improvement is ``only'' a factor 
of $\sim 30$, suggesting that some of the information added by the two features is common.
 
The inclination $\iota$ and orbital phase constant $\phi_o$
show qualitatively different behavior depending on what additional 
physics is added to the waveform model.
For both mass-ratios, we see that 
the addition of higher harmonics dramatically reduces the
long tail of large uncertainties for the parameters in the worst cases 
of the quadrupole-only results. 
The addition of the merger, on the other hand, 
results in an improvement of the most precise determinations of 
$\iota$ and $\phi_o$, with less effect on the uncertainty of the
least accurate parameter sets.
Unlike the results for luminosity distance 
and polarization phase, the overall improvement when both merger and 
harmonics are included is closer to the product of the individual 
improvements, indicating that the additional information brought by each is independent.

For the sky angles (the ecliptic latitude $\beta$ and longitude $\lambda$)
the phenomenology of the response to including higher
harmonics is roughly reversed from that seen in
the inclination and orbital phase angles.
In both Figs.~\ref{fig:histsx1} and \ref{fig:histsx2}, 
inclusion of higher harmonics most significantly improves 
the smallest uncertainties in the distribution, 
with less impact on the largest uncertainties.
The addition of the merger, however, appears to be more complicated.  
For both mass ratios the $\ell=|m|=2$ uncertainty distributions for 
both sky angles appear to uniformly improve.
In the $\ell\leq 4$ distributions, the addition of the merger
shows relatively more improvement in the least-accurate 
side of the distribution.
The reduction in uncertainty obtained by adding both features to
the waveform model shows less independence than seen with $\iota$ 
and $\phi_o$.  For the $q=1/2$ in particular, there is relatively 
little benefit to adding the higher harmonics once the merger has been included.

\subsection{Systems with different total masses}
\label{subsec:mtot}

With redshifted mass $1.33\times10^6\MSun$, the frequency of the 
inspiral-merger transition in the signals we have studied so far occurs near
the optimal region of LISA's sensitivity band.  Varying the mass shifts this 
transition frequency (in inverse proportion), thus changing LISA's relative
sensitivity to the inspiral and merger-ringdown signals. 

In Fig.~\ref{fig:histsmass}, we compare three cases, all with a mass ratio $q=1/2$ at a redshift
$z=1$, and with total masses of $M=1.33\times10^5\MSun$,
$1.33\times10^6\MSun$, and $1.33\times10^7\MSun$, chosen in part
so that the heaviest case can be compared to the results in \cite{Babak:2008bu}.
In this and subsequent figures, we do not compare results for the measurement of the mass due to the fact that our signal
duration is constant in $M$, and therefore the lightest systems do not fully span LISA's band.  This
was essentially due to computational constraints on the signal length, which we intend to improve
upon in future work.

Scaled in units of seconds, we see the best $t_c$ estimates for the mid-mass case,
which merges closest to LISA's most sensitive band, and therefore has the 
largest signal-to-noise ratio (SNR).  On the other hand, if we were to 
rescale the curves to measure
precision against the time-scale of the source physics, $M$, then the 
largest systems would be seen as most precise.

For the sky position angles the middle mass case outperforms the others by
a factor of 2-3, with a broad distribution for the highest mass case. 
For all other parameters, the lowest mass is easily the worst performer, 
with the mid-mass system marginally outperforming the largest 
mass case.  In prior investigations that were limited to the inspiral 
a more precipitous drop in performance
occurs for systems with masses approaching $10^7\MSun$.  
This is simply a result of the
absence of signal, as for such large masses the portion of the total signal that
occurs in band for LISA is increasingly dominated by the merger, so that no signal
is present when the merger is excluded.  This effect is exacerbated in studies that 
employ a more severe low-frequency cut-off in the LISA sensitivity.

\begin{figure*}
\begin{center}
\includegraphics*[trim = 0mm 0mm 0mm 8mm, clip, scale=.35, angle=0]{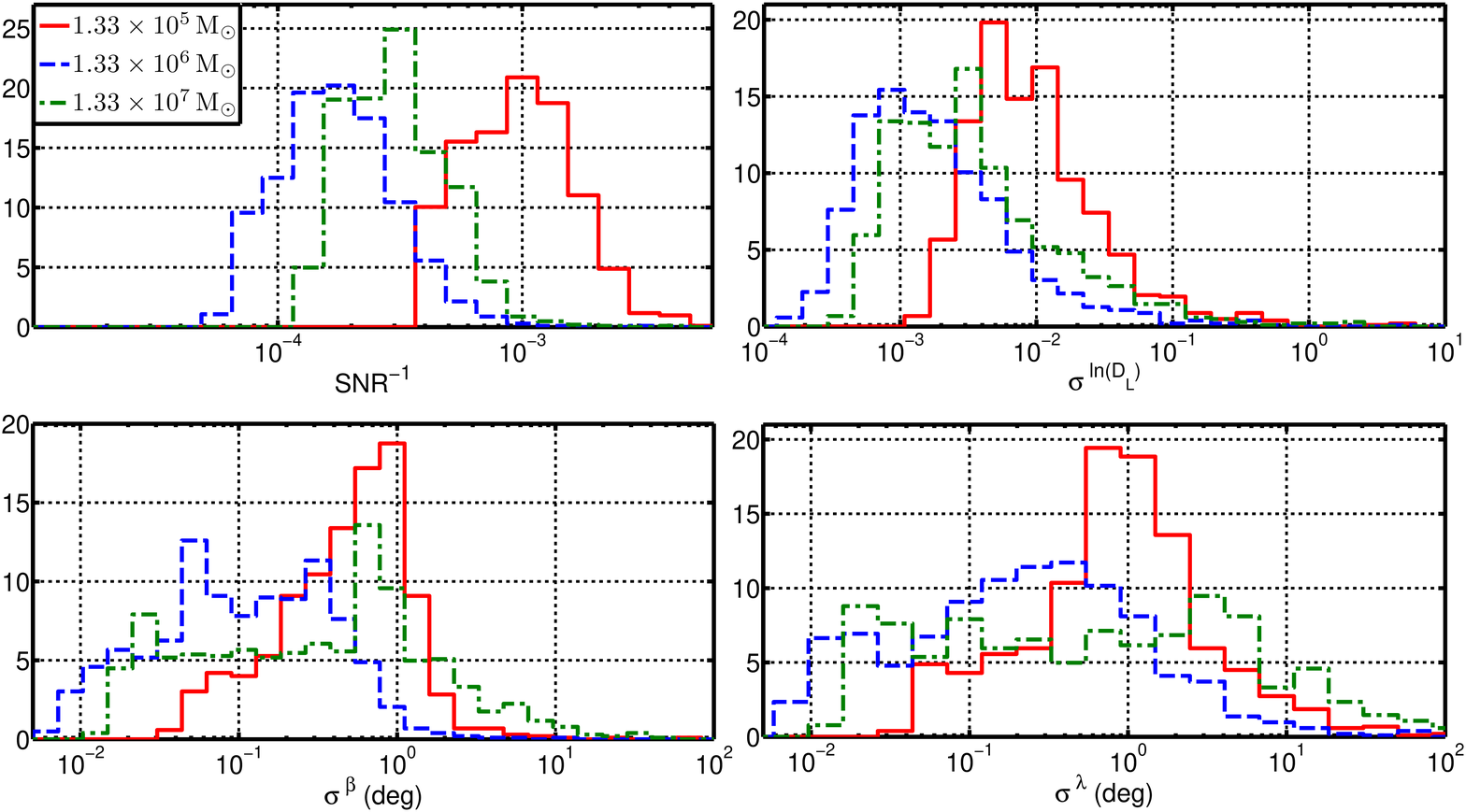}
\includegraphics*[trim = 0mm 0mm 0mm 8mm, clip, scale=.35, angle=0]{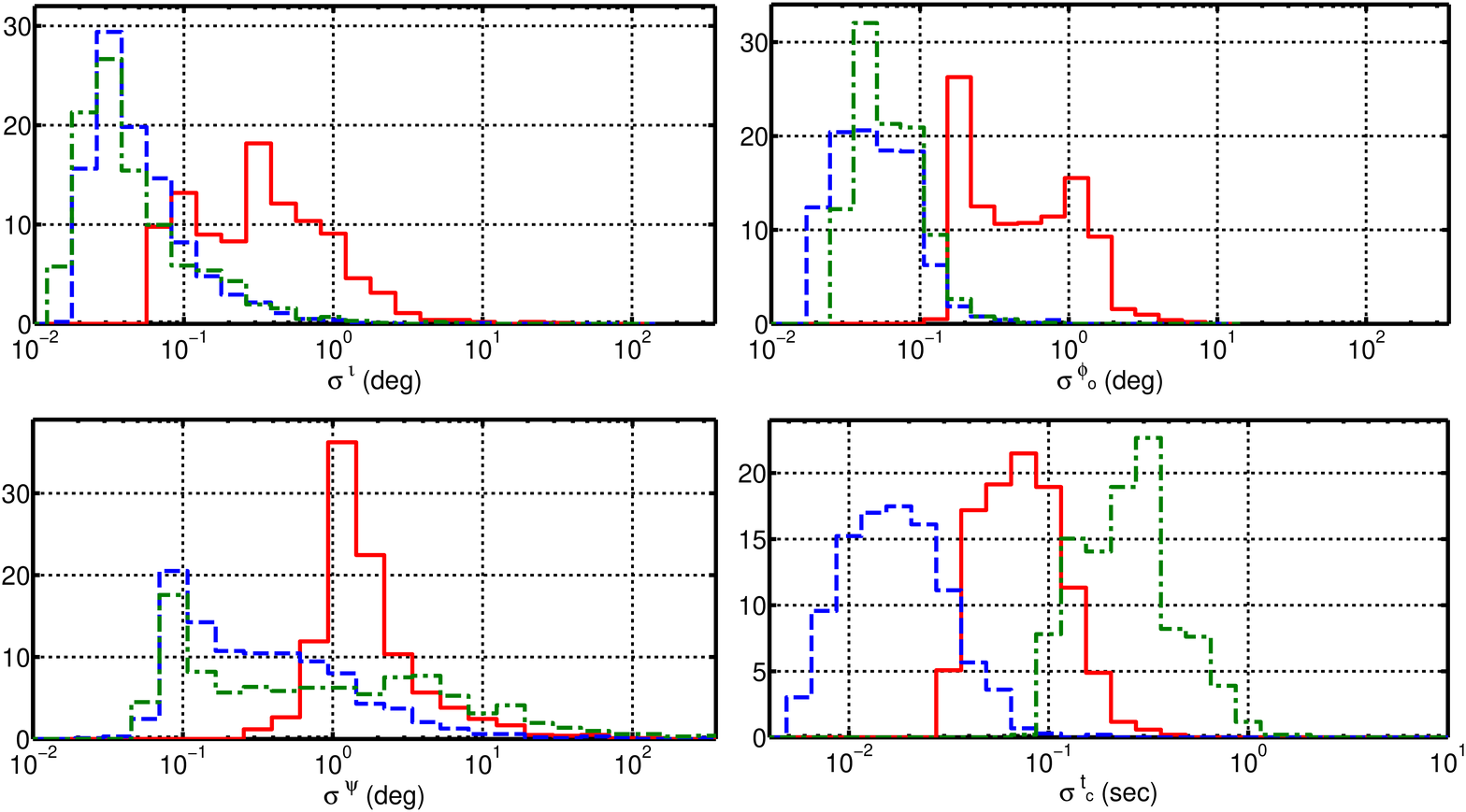}
\caption
{Uncertainty histograms for $q=1/2$ at $z=1$, corresponding to a total system mass
$M=1.33\times10^5\MSun$ (solid), $1.33\times10^6\MSun$ (dashed), and
$1.33\times10^7\MSun$ (dash-dotted).  All histogram bins are normalized by the total number of cases and are expressed as percentages.
\vspace{10mm}}
\label{fig:histsmass}
\end{center}
\end{figure*}

\subsection{Systems with different mass ratios}
\label{subsec:mrat}

We have also examined results for mass ratios other than $q=1/2$.  In
Fig.~\ref{fig:histsratio}, we compare three different mass ratios,
$q=1/2$, $q=1/4$, and $q=1/10$, with all three cases again
corresponding to a total system mass of $1.33\times10^6\MSun$ at a
redshift $z=1$.

Varying the mass ratio has surprisingly little effect on the uncertainties.
We see that the inverse SNR shows more variation than
the parameter uncertainties.  This would seem to suggest a
balance between the importance of the total signal power and the
fraction of that power contained in higher harmonics.  To the extent
that the relatively small differences among the three cases for most
parameters are statistically meaningful, the $q=1/10$ case is the
worst performer by a small margin for all parameters except $t_c$,
with an insignificant difference between the $q=1/2$ and
$q=1/4$ cases.  

\begin{figure*}
\begin{center}
\includegraphics*[trim = 0mm 0mm 0mm 8mm, clip, scale=.35, angle=0]{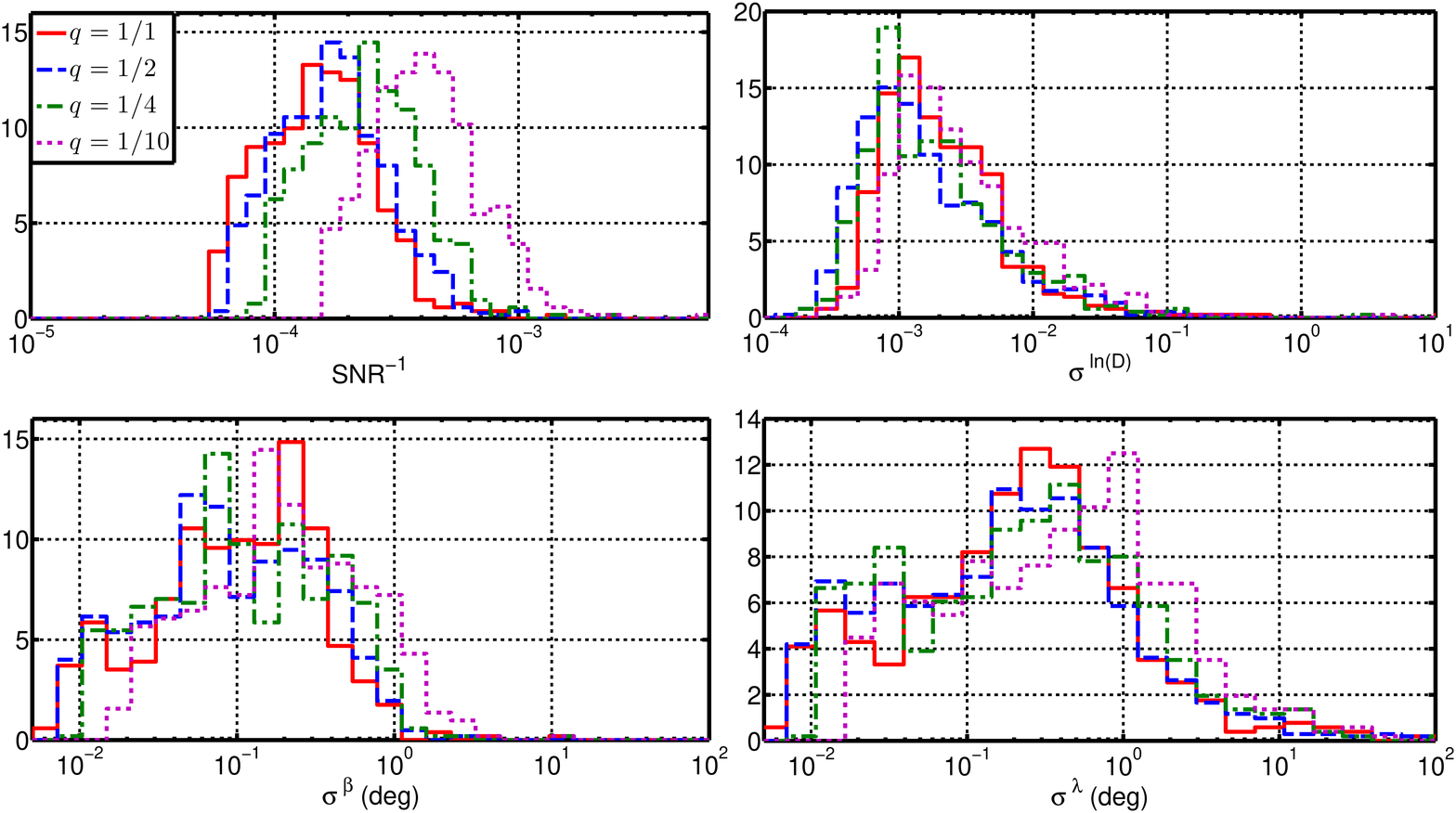}
\includegraphics*[trim = 0mm 0mm 0mm 8mm, clip, scale=.35, angle=0]{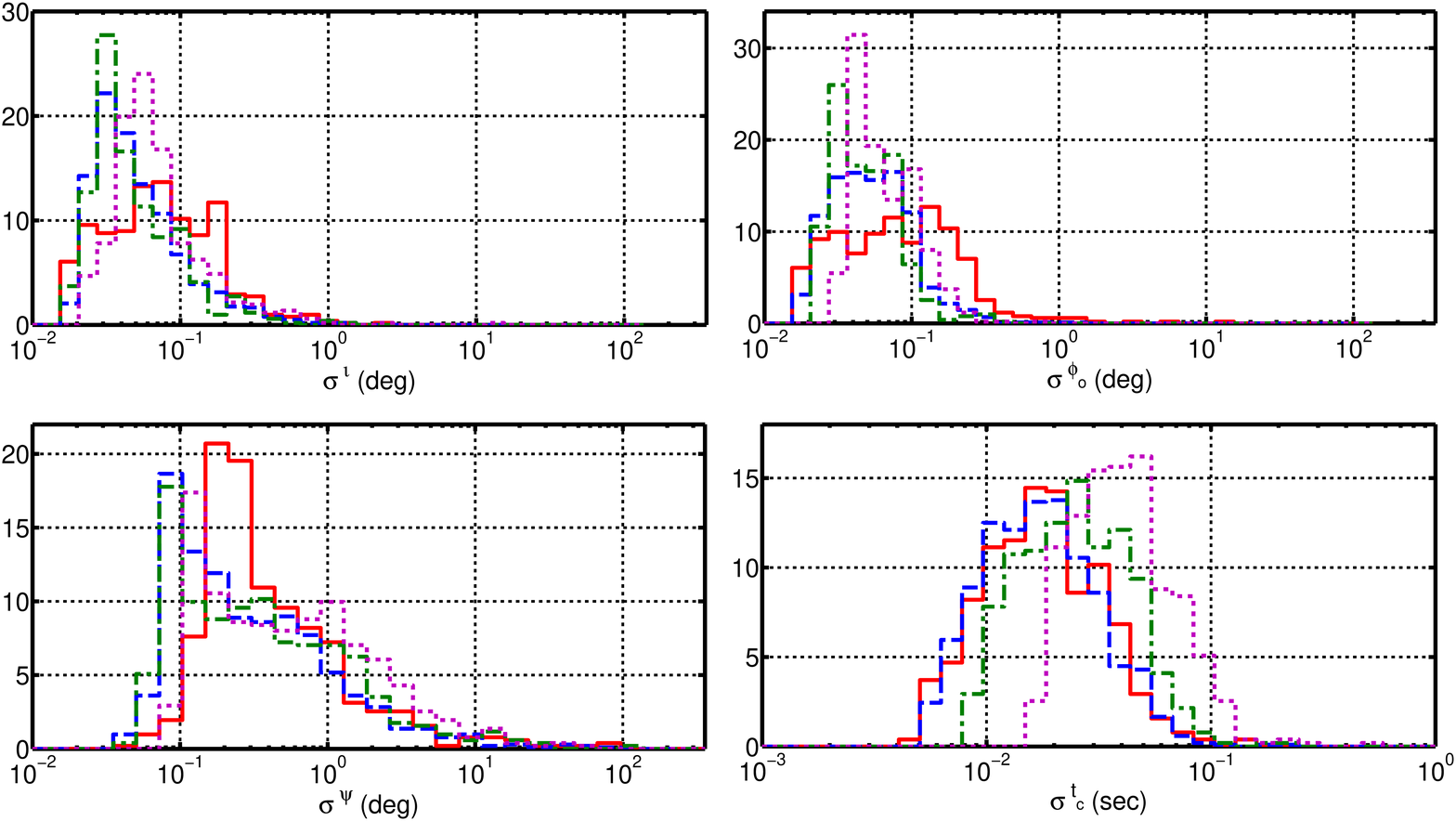}
\setcounter{figure}{5}
\caption
{Uncertainty histograms for $M=1.33\times10^6\MSun$ at $z=1$, for mass ratios
$q=1/1$ (solid), $1/2$ (dashed), and $1/4$ (dash-dotted).
All histogram bins are normalized by the total number of cases and are expressed as percentages.
\vspace{10mm}}
\label{fig:histsratio}
\end{center}
\end{figure*}

\subsection{Comparing results}
\label{subsec:comp}

In Table \ref{Table:varcomp}, we summarize our results by quoting the median parameter
uncertainties for all of our data, as well as quoting results for comparable cases from 
the literature.  We note that, as we do not include the mass ratio
in our covariance calculation, we are unable to convert to uncertainties in
the chirp mass, $\mathcal{M}_c$, and
reduced mass ratio, $\mu$, which are used in \cite{Trias:2008pu} and in most of the literature.
Furthermore, without explicit knowledge of the covariance between these parameters 
in the available publications,
we are unable to convert the results in the literature into uncertainties in the total
mass, so we leave the mass out of our comparison in Table \ref{Table:varcomp}.
We do include a comparison of the sky position, calculated using (\ref{eq:omega}).

\begin{table*}
\begin{center}
\begin{tabular}{c c c c c|c c c c c c c c c c}
\hline \hline
 & $(1+z)M$ & $m_1/m_2$ & harmonics & merger? & $\sigma^M/M$ & $\sigma^{D_L}/D_L$ & $\sigma^{\beta}$ & $\sigma^{\lambda}$ & $\Omega$ ($\mathrm{deg}^2$)& $\sigma^{\iota}$ & $\sigma^{\phi_o}$ & $\sigma^{\psi}$ & $\sigma^{t_c}$ & SNR \\
this work & 1.33e5 & 1/2 & $\ell\leq\,4$ & y & 7.0e-6 & 4.6e-3 & 0.24 & 0.39 & 2.0    & 0.26   & 0.34   & 0.84  & 4.9e-2 & 7.4e2 \\
        - & 1.33e6 & 1/1 & $\ell\leq\,4$ & y & 6.5e-6 & 1.9e-3 & 0.12 & 0.22 & 0.11   & 7.8e-2 & 7.1e-2 & 0.34 & 1.8e-2 & 4.7e3\\
        - & 1.33e6 & 1/2 & $\ell\leq\,4$ & y & 7.7e-6 & 1.3e-3 & 8.5e-2 &0.18& 5.6e-2 & 4.2e-2 & 5.1e-2 & 0.23 & 1.7e-2 & 4.2e3 \\
        - & 1.33e6 & 1/4 & $\ell\leq\,4$ & y & 8.8e-6 & 1.3e-3 & 0.10 & 0.24 & 9.0e-2 & 4.6e-2 & 3.9e-2 & 0.28 & 4.4e-2 & 3.8e3\\
        - & 1.33e6 & 1/10& $\ell\leq\,4$ & y & 4.7e-6 & 2.1e-3 & 0.19 & 0.39 & 0.28   & 6.0e-2 & 5.8e-2 & 0.43 & 4.4e-2 & 1.7e3 \\
        - & 1.33e7 & 1/2 & $\ell\leq\,4$ & y & 1.1e-5 & 2.9e-3 & 0.37 & 0.61 & 0.64   & 5.8e-2 & 3.6e-2 & 0.70 & 0.32   & 2.5e3 \\
\hline
        - & 1.33e6 & 1/1 & $\ell\leq\,4$ & n & 7.6e-6 & 6.8e-3 & 0.57 & 1.4  & 3.0    & 0.18   & 0.19   & 1.7  & 0.23   & 1.5e3 \\
        - & 1.33e6 & 1/1 & $\ell=2$      & y & 6.5e-6 & 4.4e-3 & 0.22 & 0.33 & 0.30   & 0.14   & 0.12   & 0.86 & 2.1e-2 & 4.7e3 \\
        - & 1.33e6 & 1/1 & $\ell=2$      & n & 7.6e-6 & 6.2e-2 & 3.5  & 5.6  & 83.    & 1.5    & 1.3    & 12.  & 0.27   & 1.5e3 \\
        - & 1.33e6 & 1/2 & $\ell\leq\,4$ & n & 9.3e-6 & 4.6e-3 & 0.47 & 1.0  & 1.6    & 0.12   & 0.14   & 1.1  & 0.25   & 1.3e3 \\
        - & 1.33e6 & 1/2 & $\ell=2$      & y & 7.8e-6 & 5.9e-3 & 0.26 & 0.41 & 0.46   & 0.16   & 0.20   & 1.1  & 2.2e-2 & 4.2e3 \\
        - & 1.33e6 & 1/2 & $\ell=2$      & n & 9.6e-6 & 7.5e-2 & 3.8  & 6.1  & 100.   & 1.9    & 2.4    & 15.  & 0.33   & 1.3e3 \\
- & \footnote{signal duration is limited to 10 cycles, for comparison to \cite{Babak:2008bu}}
            1.33e7 & 1/2 & $\ell\leq\,4$ & y & 8.7e-5 & 2.9e-3 & 0.39 & 0.63 & 0.75   & 6.9e-2 & 3.7e-2 & 0.70 & 0.73   & 2.5e3 \\
\hline
\cite{Babak:2008bu} & 
            1.33e7 & 1/2 & $\ell\leq\,4$ & y & -      & -      &4.6e-2&5.7e-2& - & - & - & - & - & - \\
\cite{Trias:2008pu} & 
               4e6 & 1/1 & 2.5 PN        & n & -      & 2.5e-2 & -    & -    & 4.2    & - & - & -              & 25.    & - \\
        - &  2.2e6 & 1/10 & 2.5 PN       & n & -      & 1.7e-2 & -    & -    & 2.6    & - & - & -              & 13.    & - \\
\footnote{estimated from histograms in Fig.~2 of \cite{Vecchio:2003tn}}\cite{Vecchio:2003tn} 
             & 4e6 & 1/1 & \footnote{``R'' indicates that the amplitude was restricted to the leading order term.}
                           R1.5 PN       & n & -      & 2e-2   & -   & -     & 0.6    & - & - & - & - & - \\
\footnote{estimated from the results in Table II of \cite{Cutler:1997ta}}\cite{Cutler:1997ta} 
             & 4e6 & 1/1 & R1.5 PN       & n & -      & 7e-2   & -   & -     & 0.5    & - & - & - & - & - \\
\hline \hline
\end{tabular}
\end{center}
\caption
{Median fractional variance of all the extrinsic parameters for the cases
investigated in this paper, as well as results from the
literature for comparable systems.  All angles are measured in degrees, and time is measured in seconds.  
We separate our results into cases where all available information has been included (top portion), and where
some information has been suppressed for testing and analysis (bottom portion).  For literature results, ``$X$ PN'' refers 
to the post-Newtonian order of the model used.
All studies set at a fixed source distance of $z=1$, albeit with slightly different cosmological
parameters.  The results in this work correspond to $\sim10^6\,M$ of observation,
($\sim 3$ months for $M=1.33\times10^6\,\MSun$) unless otherwise
noted, while the results from the literature correspond to 1 year of observation.
\vspace{10mm}}
\label{Table:varcomp}
\end{table*}

Of particular note is the discrepancy between our results, and the results found in \cite{Babak:2008bu}.  Specifically, from their Fig.~ 1,
their median latitude and longitude uncertainties are 0.046 deg and 0.057 deg, respectively, or $\sim 3$ arcmin as stated in their abstract.
However, when running as identical a case as possible with ten cycles prior to merger of a system with $q=1/2$ and $M=1.33\times10^7 {\rm M}_{\odot}$,
and using the $\Abar$ and $\Ebar$ channels only, we arrive at median estimates of 0.39 deg and 0.63 deg for the latitude and longitude, respectively.
This represents an order-of-magnitude disagreement.  
We note that for a total mass of $M=1.33\times10^6 {\rm M}_{\odot}$,
the median latitude and longitude uncertainties for all mass ratios was within a factor of a few of the localization claimed in \cite{Babak:2008bu},
with $q=1/2$ providing the best localization with median latitude and longitude uncertainties of 0.09 deg (5 arcmin) and 0.18 deg (11 arcmin), respectively, and with 10\% of the cases in that ensemble
being localized at the $\sim 1$ arcmin level.

\subsection{Accumulation of information with time}
\label{subsec:timeEvo}

In actual observations of black hole binaries with LISA, information about 
the system will be progressively unveiled over time.  In particular some
estimate of the system parameters may be available in advance of the merger
observation.  As the system approaches merger the uncertainties of these 
estimates are expected to decrease sharply \cite{Lang:2006bz}.  This real-time
development is especially important in planning multi-messenger observations.
How and when sky position estimates improve as the coalescence proceeds may
impact the instruments operational requirements 
including how frequently data downlinks are 
required and in planning protected observing periods near the moment of 
merger.  In turn, these operational requirements may influence details of LISA's 
instrumental design.

In order to compare the evolution in measured parameter precision for 
different systems, 
we have calculated the parameter uncertainty
for waveforms whose ends are ``turned off'' via windowing as described 
earlier.  In this section, all the specified times correspond to 
the time at the mid-point of the applied taper.  
This procedure is analogous to a realistic
procedure for measuring parameters from progressively longer segments of 
real-time data.

In Fig.~\ref{fig:x1x2tcut}, 
we compare the uncertainty in ecliptic latitude $\beta$ and longitude $\lambda$
for the equal-mass waveform and the $q=1/2$ waveform, both with higher 
harmonics $\ell \leq 4$ and restricted to quadrupolar modes ($\ell=2, m=\pm 2$).
The linear-appearing decrease seen in both panels indicates that, 
over the last several hours before merger, our estimates for uncertainties in the
sky position angles are roughly proportional to the time remaining 
before merger until a couple minutes before ``merger'' 
(which we have defined as the moment at which the $|\ell|=2$ mode 
amplitude peaks).  For the cases studied, the dominant ringdown 
radiation period is about $80$ s, roughly setting the scale at which the 
linear trend levels off.  Note that, in some cases the parameter 
uncertainties may continue to improve after ``merger'', drawing on information
in the ringdown radiation.  The lower pair of curves in each panel are based 
on the waveform model including the higher harmonics.  Consistent with the
discussion in Sec. \ref{subsec:harm}, including the harmonics continues to be valuable
even late in the observation, after the merger is recorded.

\vspace{10mm}
\begin{figure*}
\includegraphics*[trim = 0mm 0mm 0mm 0mm, clip, scale=.16, angle=0]{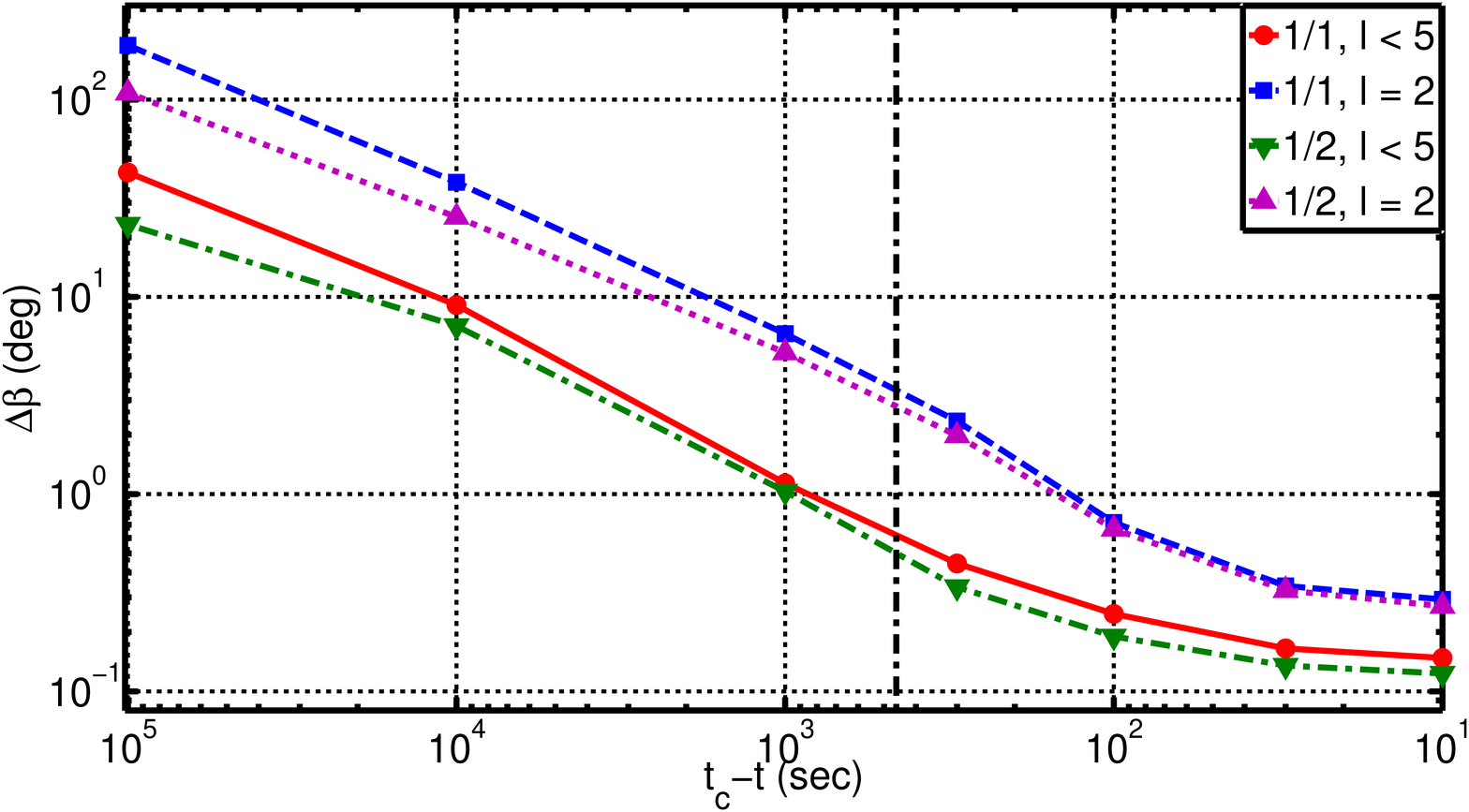}
\includegraphics*[trim = 0mm 0mm 0mm 0mm, clip, scale=.16, angle=0]{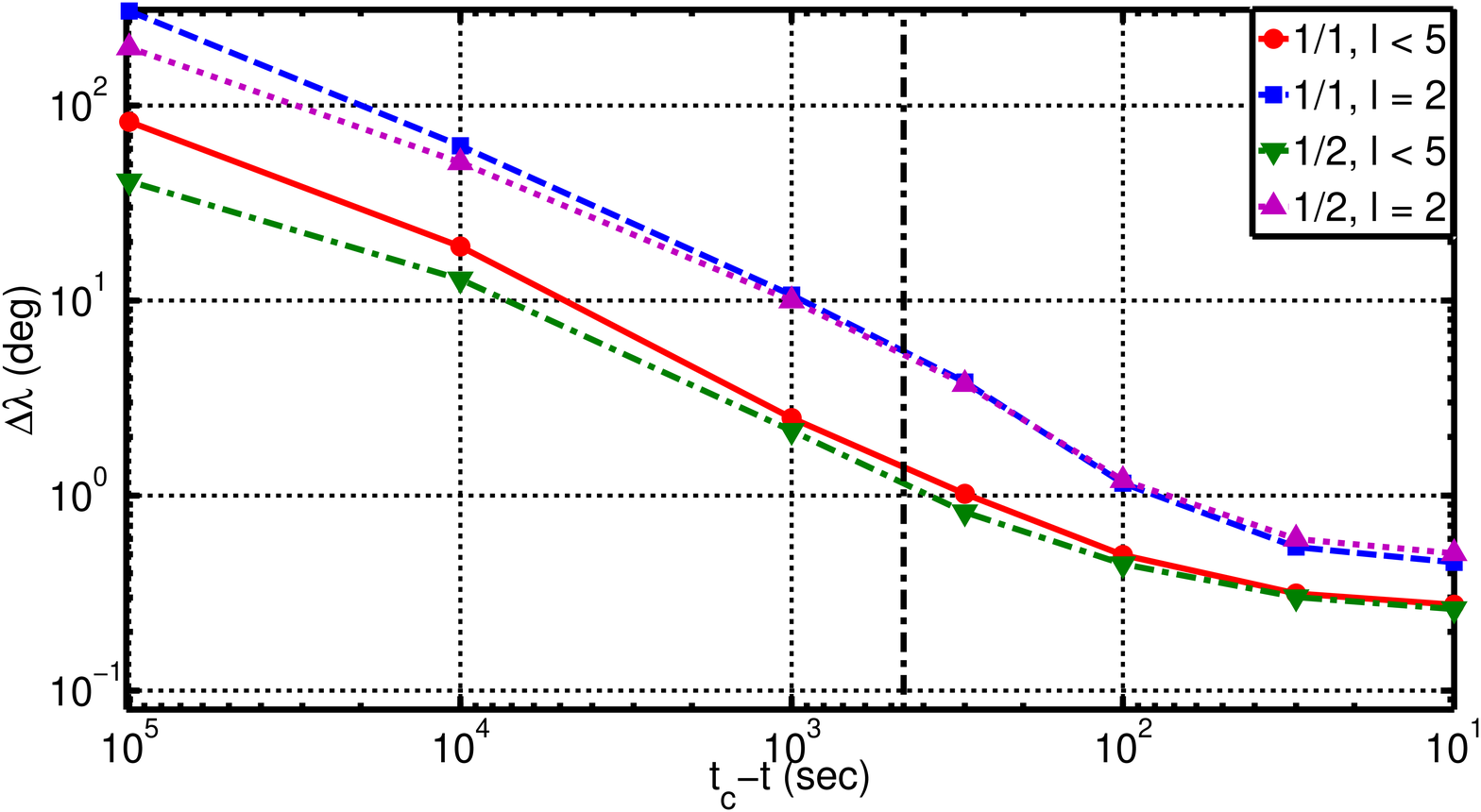}
\setcounter{figure}{6}
\caption
{Comparison of the median ecliptic latitude (left panel) and 
longitude (right panel) uncertainties as a function of time
before the merger for a mass $M=1.33\times10^6\MSun$
at $z=1$, for mass ratios of $q=1$ and $q=1/2$.
For each mass ratio, we compare the uncertainty for a
waveform including harmonics $\ell \leq 4$ ($q=1$: solid, circles; $q=1/2$: dash-dotted,
down arrows), and for waveforms including only the quadrupole ($\ell=2$, $m=\pm 2$) modes
($q=1$: dashed, squares; $q=1/2$: dotted, up arrows).
The vertical dash-dotted line corresponds
to the Schwarzschild ISCO.\vspace{10mm}}\label{fig:x1x2tcut}
\end{figure*}

\begin{figure*}
\includegraphics[trim = 0mm 0mm 0mm 0mm, clip, scale=.16, angle=0]{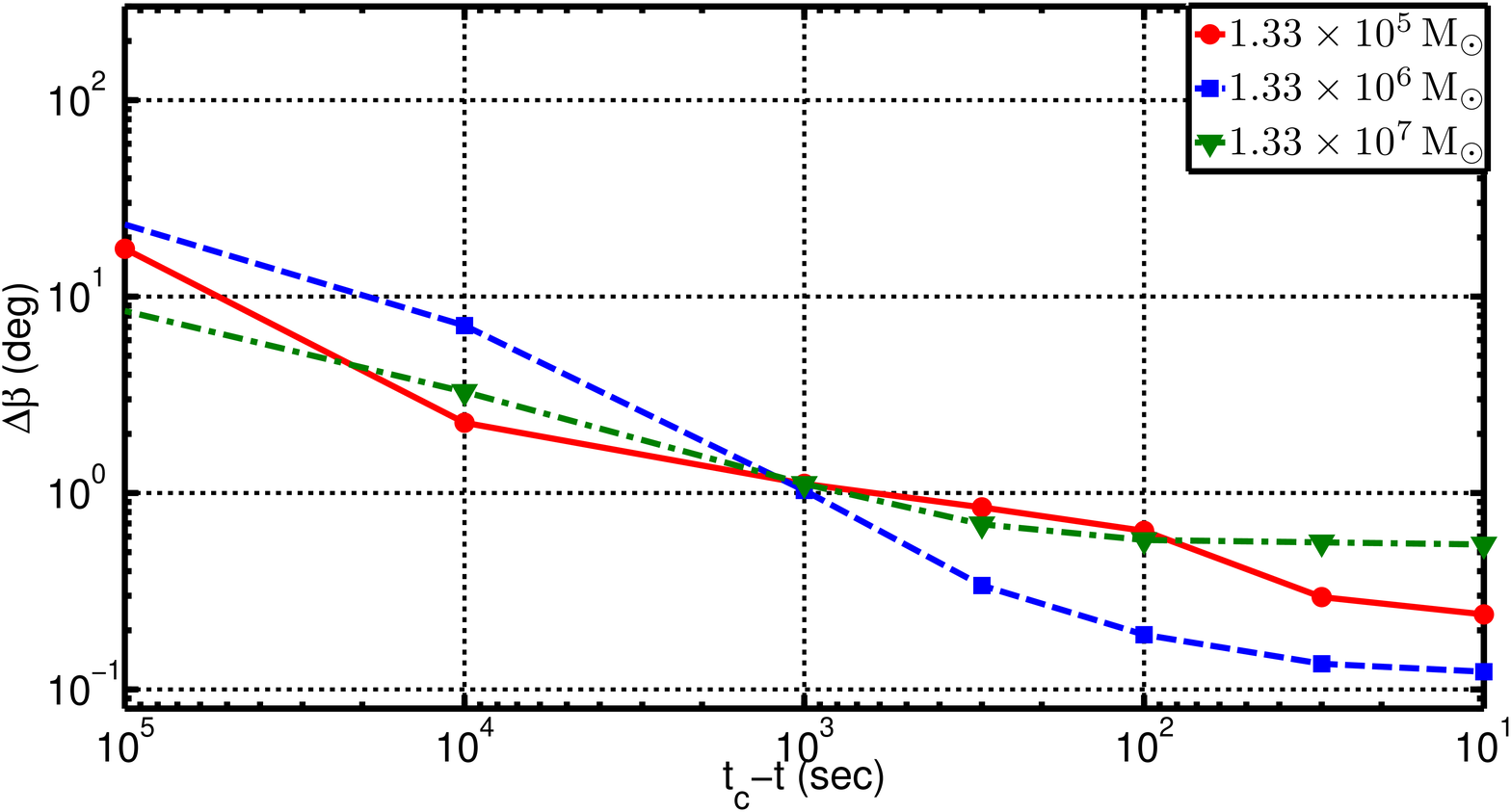}
\includegraphics[trim = 0mm 0mm 0mm 0mm, clip, scale=.16, angle=0]{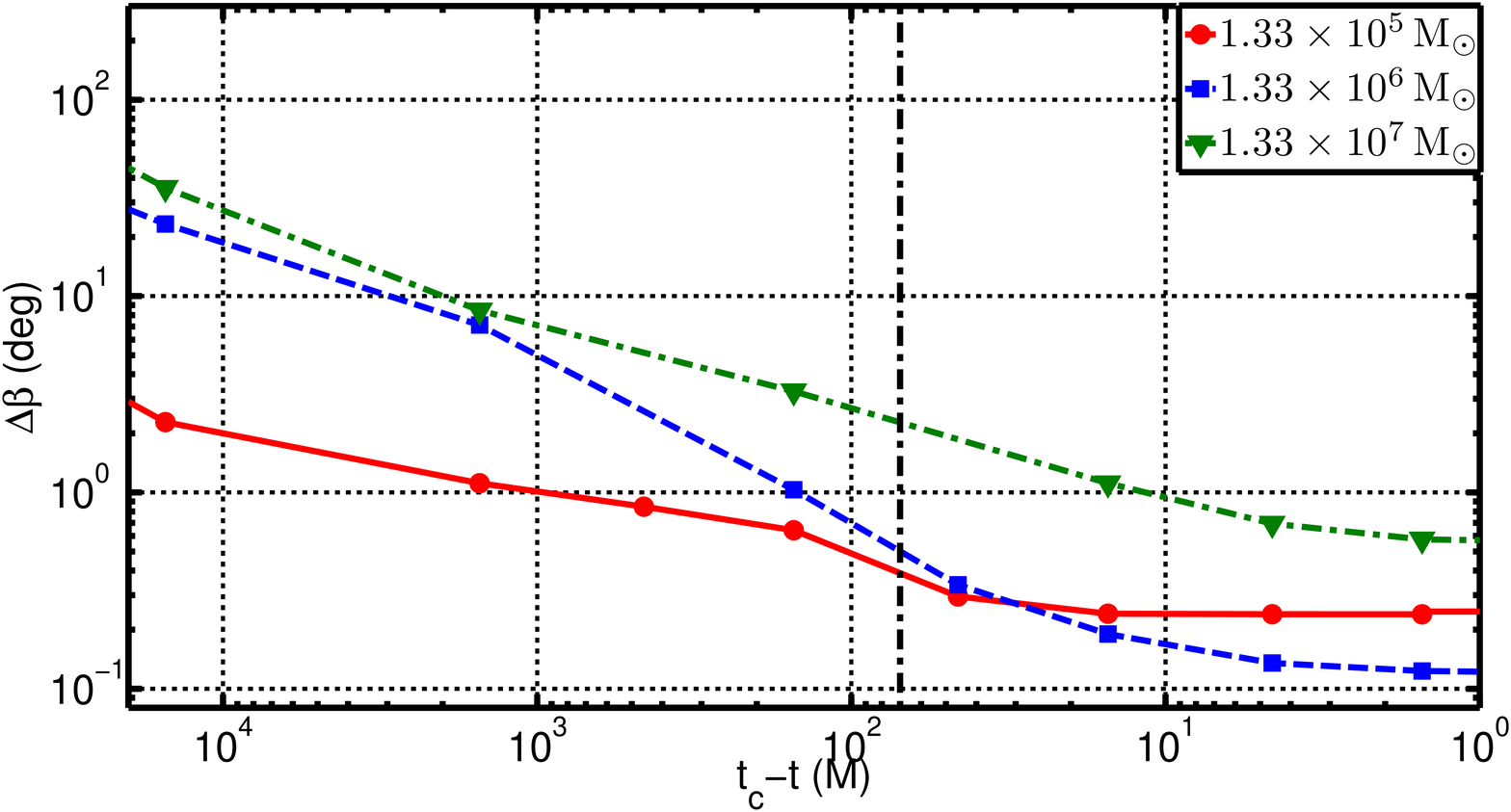}
\caption
{Comparison of the latitude uncertainty as a function of time
before the merger for a mass ratio $q=1/2$ at $z=1$, with a total system mass  
$M=1.33\times10^5\MSun$ (solid, circles), 
$M=1.33\times10^6\MSun$ (dashed, squares), and
$M=1.33\times10^7\MSun$ (dash-dotted, down arrows).  
In the right panel, the time axis is scaled in units of $M$, 
the physical time-scale of the merger process, 
rather than seconds.  The vertical dash-dotted line corresponds
to the Schwarzschild ISCO. \vspace{10mm}}\label{fig:e5e6e7tcut}
\end{figure*}

Fig.~\ref{fig:e5e6e7tcut} shows another comparison of latitude uncertainty, 
for the same systems compared in Fig.~\ref{fig:histsmass}.  Because 
the parameter being varied is the total system mass, and the mass 
rescales time, we compare these results using
times measured in $M$ and in seconds.
Because these signals are simply 
mass-rescalings of the same signal in naturalized units, and therefore have identical
harmonic content relative to their quadrupolar content, the main factor for 
differences is the frequency band spanned by the signal, and where that band falls 
relative to the most sensitive band of the detector.  The lowest-mass case, 
$M=1.33\times10^5\MSun$,
has the largest number of cycles in-band, and therefore performs best at early 
times.  By the time it merges, however, the signal is chirping at frequencies 
much higher than the most sensitive band for LISA, so the the contribution after ISCO is
negligible for this case.  The mid-mass case, $M=1.33\times10^6\MSun$, 
is outperformed by the lowest-mass case at early times.  However, because it merges 
in LISA's most sensitive band, the contribution approaching ISCO and running 
through the merger
and ringdown is far greater than the other cases.  Indeed, by the time the full 
signal has been included, this case yields a more accurate estimate than the 
lowest-mass case by a factor of $\sim2$.  The largest mass, $M=1.33\times10^7\MSun$,
has the fewest cycles in band, so it yields the lowest precision at early 
times.  However, it too has a substantial gain in SNR, relative to the SNR of its 
inspiral, in the late inspiral through the merger and ringdown, so it too makes 
gains in precision
relative to the lowest-mass case, although unlike the $M=1.33\times10^6\MSun$ 
case, it does not fully ``catch up'', and remains the worst performer of the three.
 
\vspace{10mm}

\begin{figure}
\includegraphics[trim = 0mm 0mm 0mm 0mm, clip, scale=.16, angle=0]{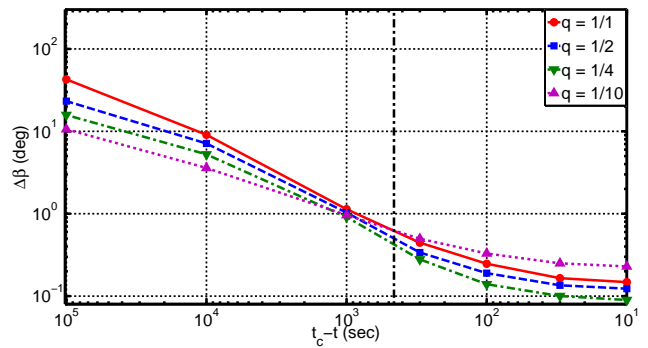}
\caption
{Comparison of the latitude uncertainty as a function of time
before the merger for a mass $M=1.33\times10^6\MSun$
at $z=1$, and mass ratios of $q=1$ (solid, circles), 
$q=1/2$ (dashed, squares), $q=1/4$ (dotted, down arrows), and $q=1/10$ (dash-dotted, up arrows).
The vertical dash-dotted line corresponds
to the Schwarzschild ISCO frequency. \vspace{10mm}}\label{fig:mrattcut}
\end{figure}

We compare latitude uncertainty for four different mass ratios in
Fig.~\ref{fig:mrattcut}: $q=1$, $q=1/2$, $q=1/4$, and $q=1/10$.  This comparison again shows a 
trade-off between the number of in-band cycles and the signal power.  
Because radiation reaction is weaker for more disparate mass ratios, 
the $q=1/10$ yields the highest precision at early times, despite 
having significantly less power (the SNR scales as $\eta$ for the 
inspiral, and as $\eta^2$ for the merger 
\cite{Flanagan:1997sx, McWilliams_PhD}).  
Sky position uncertainty for the smaller-$q$ cases decreases 
more slowly over most of the last day than the near-linear rate
seen for the equal-mass case.
By $\sim 20$ minutes before merger, approaching ISCO, 
the median uncertainties in $\beta$
are roughly the same for all mass-ratios shown. 
At late times, the power content 
becomes a more dominant factor in further decreasing the uncertainty in
$\beta$. The equal-mass case contains more signal power in the merger.  
The $q=1/2$ and $q=1/4$ cases are nearly optimal, retaining some of the merger
signal strength but perhaps benefiting more from stronger harmonic 
content at late times \cite{Baker:2008mj}.  Overall final sky-position error estimates
are nearly flat for $1>q>1/4$ (see Table \ref{Table:varcomp}).  
By $q=1/10$ the merger signal power is significantly diminished.  While there
are still improvements in position estimates after merger, they are notably
smaller than those in the other cases.

\vspace{10mm}

\section{Conclusion}
\label{sec:conc}

We have investigated the precision with which black hole binary system
parameters can be measured from LISA observations
including merger waveforms in the analysis of nonspinning binaries
with moderate mass ratios ($q\geq 1/10$).  
We have further studied how the
expected performance depends on mass ratio and total
system mass, and the impact of including or neglecting
the merger signal and higher harmonics.  
The luminosity distance 
and the polarization phase depend on both the inclusion of the merger
and the presence of higher harmonics, although the improvements from including
these two elements are not independent.
The inclination, the orbital phase
constant, the ecliptic latitude and ecliptic longitude
also depend on both the merger and the harmonic content. For these parameters
the improvements resulting from including both the merger and higher harmonics 
are essentially independent.

For comparable-mass 
systems near $10^6\MSun$, ignoring the merger reduces the SNR by a factor of
$\sim 3$ and results in a similar loss of median precision in parameter estimation, 
even more so for the sky position estimates.  
For sky position, ignoring the merger results in a more significant loss of precision than ignoring higher harmonics.  
Parameter estimates are roughly independent of mass ratio
through $1>q>1/4$, for sky position in particular, though for smaller mass ratios $q\lesssim1/10$ the 
precision begins to decrease.  For  $q=1/2$, 
the best parameter estimates are obtained for systems
near $10^6\MSun$, which merge in the middle of LISA's sensitivity band.  
Decreasing the mass by an order of magnitude to $\sim 10^5\MSun$ results in a precision loss of
roughly a factor of 5 with a diminished relative contribution from the merger.  Increasing the mass to  
$\sim 10^7\MSun$ results in a similar loss.  
Though we have left out the effects of spin, including precession, 
our median sky position precision estimates are similar to those obtained with 
precession, but ignoring the merger \cite{Lang:2006bz}. 
Each method locates the systems 
in the sky within ${\cal O}(10$ arcmin).  
Our best cases ($\sim$ top 10\%) are localized 
 at the level of ${\cal O}(1$ arcmin).  
We estimate that LISA will usually be able to locate
larger mass systems (near $10^7\MSun$) quite well, 
in some cases better than systems with masses near $10^5\MSun$, and
far better than earlier estimates based on inspirals alone.  
Our results for these more massive systems do not, however, reproduce the preliminary 
(but widely discussed) extraordinarily precise sky localization results 
found in \cite{Babak:2008bu}, though we do achieve such high precision for the $\sim10^6\MSun$ systems, where both
the inspiral and merger are in-band and can contribute.
For equal-mass systems near $10^6\MSun$ the sky angle estimates improve over the 
last several hours up to a few minutes before merger in rough proportion to the time remaining before merger.

\acknowledgments

We thank Alessandra Buonanno and Ryan Lang for thorough reviews of the manuscript, 
and Keith Arnaud and Tuck Stebbins for useful discussions.  STM was supported by an
appointment to the NASA Postdoctoral Program at the Goddard Space
Flight Center, administered by Oak Ridge Associated Universities through
a contract with NASA.  The simulations were carried out using resources from
the NASA Center for Computational Sciences (Goddard Space Flight Center).

\end{document}